\documentclass[12pt]{article}
\pdfoutput=1

\usepackage[usenames,dvipsnames,svgnames,table]{xcolor} 
\usepackage[obeyspaces,hyphens,spaces]{url}
\usepackage{jcapmod}
\usepackage{verbatim}

\usepackage{booktabs}
\usepackage[english]{babel}
\usepackage{amsmath, amssymb, amsbsy, amstext, amsthm, simplewick}
\usepackage{hyperref}
\usepackage{graphicx}
\usepackage{amsfonts}
\usepackage{upgreek}
\usepackage{framed}
\usepackage{tensor}
\usepackage{pifont}
\usepackage{latexsym, mathrsfs}
\usepackage{array}
\usepackage{hyperref}
\usepackage{xspace}
\usepackage{longtable}
\usepackage{multirow}
\usepackage{cases}
\usepackage{empheq}

\usepackage{bm}
\usepackage{bbm}
\usepackage{afterpage}
\usepackage{setspace}
\usepackage{braket}

\usepackage[load-configurations=astronomy, range-units=brackets, range-phrase=-, per-mode=reciprocal, mode=math]{siunitx}
\usepackage{threeparttable}
\usepackage{subfig}

\usepackage{ragged2e}

\usepackage{hhline}
\usepackage{array}
\newcolumntype{P}[1]{>{\centering\arraybackslash}p{#1}}
\usepackage{booktabs}
\usepackage{tikz}
\usetikzlibrary{calc,shadings,patterns,tikzmark,fadings}
\definecolor{Blue}{rgb}{0.25, 0.41, 0.88}
\definecolor{Red}{rgb}{0.92,0.,0.}
\definecolor{darkorange}{rgb}{1.0,0.549,0.}
\definecolor{cobalt}{RGB}{44, 98, 120}
\definecolor{Mathematica1}{rgb}{0.368417, 0.506779, 0.709798}
\definecolor{Mathematica2}{rgb}{0.880722, 0.611041, 0.142051}
\definecolor{Mathematica3}{rgb}{0.560181, 0.691569, 0.194885}
\definecolor{Mathematica4}{rgb}{0.922526, 0.385626, 0.209179}
\definecolor{Mathematica5}{rgb}{0.528488, 0.470624, 0.701351}
\definecolor{Mathematica6}{rgb}{0.772079, 0.431554, 0.102387}
\definecolor{Mathematica7}{rgb}{0.363898, 0.618501, 0.782349}
\definecolor{Mathematica8}{rgb}{1, 0.75, 0}
\definecolor{Mathematica9}{rgb}{0.647624, 0.37816, 0.614037}
\definecolor{plotBlue}{RGB}{94, 130, 181}
\definecolor{plotRed}{RGB}{233, 85, 54}
\definecolor{plotGreen}{RGB}{142, 176, 50}
\definecolor{plotPurple}{RGB}{135, 120, 178}

\newcolumntype{C}[1]{>{\centering\let\newline\\\arraybackslash\hspace{0pt}}m{#1}}

\makeatletter
\newlength{\apb@width}
\newcommand{\autoparbox}[2][c]{\settowidth{\apb@width}{#2}\parbox[#1]{\apb@width}{#2}}

\makeatother

\makeatletter
\newsavebox\myboxA
\newsavebox\myboxB
\newlength\mylenA

\newcommand*\xoverline[2][0.75]{
    \sbox{\myboxA}{$\m@th#2$}%
    \setbox\myboxB\null
    \ht\myboxB=\ht\myboxA%
    \dp\myboxB=\dp\myboxA%
    \wd\myboxB=#1\wd\myboxA
    \sbox\myboxB{$\m@th\overline{\copy\myboxB}$}
    \setlength\mylenA{\the\wd\myboxA}
    \addtolength\mylenA{-\the\wd\myboxB}%
    \ifdim\wd\myboxB<\wd\myboxA%
       \rlap{\hskip 0.5\mylenA\usebox\myboxB}{\usebox\myboxA}%
    \else
        \hskip -0.5\mylenA\rlap{\usebox\myboxA}{\hskip 0.5\mylenA\usebox\myboxB}%
    \fi}
\makeatother

\numberwithin{equation}{section}

\def\beq{\begin{equation}}
\def\eeq{\end{equation}}

\def\bea{\begin{eqnarray}}
\def\eea{\end{eqnarray}}

 \def\be{\begin{equation}}
 \def\ee{\end{equation}}
 \def\bes{\begin{eqnarray}}
 \def\ees{\end{eqnarray}}

\def\beq{\begin{equation}}
\def\eeq{\end{equation}}
\def\bea{\begin{eqnarray}}
\def\eea{\end{eqnarray}}

\DeclareRobustCommand{\SkipTocEntry}[4]{}

\usepackage{colortbl}
\definecolor{blue2}{cmyk}{1, 0.1, 0.1, 0.1}

\definecolor{pyBlue}{RGB}{31, 119, 180}
\definecolor{pyRed}{RGB}{214, 39, 40}
\definecolor{pyGreen}{RGB}{44, 160, 44}
\definecolor{pyBlue2}{RGB}{0, 111, 237}
\definecolor{pyRed2}{RGB}{224, 52, 36}

\begin{document}

\pagenumbering{roman}
\begin{titlepage}
\baselineskip=10.5pt \thispagestyle{empty}

\bigskip\

\vspace{0cm}
\begin{center}
	~{\Huge \textcolor{Sepia}{\bf \sffamily ${\cal C}$}}{\large \textcolor{Sepia}{\bf\sffamily HAO${\cal S}$}} ~{\Huge 
	 ~{\large \textcolor{Sepia}{\bf\sffamily ${\cal A}$N${\cal D}$}}~ {\Huge \textcolor{Sepia}{\bf\sffamily ${\cal C}$}}{\large \textcolor{Sepia}{\bf\sffamily OMPLEXIT${\cal Y}$}}\vspace{0.15cm} \\{\large \textcolor{Sepia}{\bf\sffamily ${\cal F}$}}{\large \textcolor{Sepia}{\bf\sffamily RO${\cal M}$}}\vspace{0.45cm}~\\ {\Huge \textcolor{Sepia}{\bf\sffamily ${\cal Q}$}}{\large \textcolor{Sepia}{\bf\sffamily UANTU${\cal M}$}}~
	{\Huge \textcolor{Sepia}{\bf\sffamily ${\cal N}$}}{\large \textcolor{Sepia}{\bf\sffamily EURA${\cal L}$}}~ {\Huge \textcolor{Sepia}{\bf\sffamily ${\cal N}$}}{\large \textcolor{Sepia}{\bf\sffamily ETWOR${\cal K}$}}}\\  \vspace{0.25cm}
	{\fontsize{13}{16}\selectfont  \bfseries \textcolor{Sepia}{A study with Diffusion Metric in Machine Learning}}
\end{center}

\vspace{0.05cm}
		\begin{center}
	{\fontsize{17}{18}\selectfont Sayantan Choudhury${}^{\textcolor{Sepia}{1,2,3},}$\footnote{{\sffamily \textit{ Corresponding author, E-mail}} : {\ttfamily sayanphysicsisi@gmail.com}}}${{}^{,}}$
	\footnote{{\sffamily \textit{ NOTE: This project is the part of the non-profit virtual international research consortium ``Quantum Aspects of Space-Time \& Matter" (QASTM)} }. }${{}^{,}}$ \\
	{\fontsize{17}{18}\selectfont Ankan Dutta${}^{\textcolor{Sepia}{4}}$}~ and
	~{\fontsize{17}{18}\selectfont Debisree Ray${}^{\textcolor{Sepia}{5}}$}~

\end{center}

\begin{center}
	\vskip4pt
	{ 
		\textit{${}^{1}$National Institute of Science Education and Research, Bhubaneswar, Odisha - 752050, India}\\
		\textit{${}^{2}$Homi Bhabha National Institute, Training School Complex, Anushakti Nagar, Mumbai - 400085, India}\\
			\textit{${}^{3}$ Max Planck Institute for Gravitational Physics (Albert Einstein Institute),\\
			Am M$\ddot{u}$hlenberg 1,
			14476 Potsdam-Golm, Germany.}\\
		\textit{${}^{4}$Department of Mechanical Engineering, Jadavpur University, Kolkata - 700032, India}\\
		\textit{${}^{5}$ Department of Physics and Astronomy, Mississippi State University, 355 Lee Boulevard
Mississippi State, MS 39762, United States}\\
	}
\end{center}

\vspace{1.0cm}
\hrule \vspace{0.3cm}
~~~~~~~~~~~~~~~~~~\quad\quad\quad~~~~~~~~~~~~\quad\quad~~~\noindent {\bf Abstract}\\[0.1cm]

  In this work, our prime objective is to study the phenomena of quantum chaos and complexity in the machine learning dynamics of {\it Quantum Neural Network} (QNN). A {\it Parameterized Quantum Circuits} (PQCs) in the hybrid quantum-classical framework is introduced as a universal function approximator to perform optimization with {\it Stochastic Gradient Descent} (SGD). We employ a statistical and differential geometric approach to study the learning theory of QNN. The evolution of parametrized unitary operators is correlated with the trajectory of parameters in the Diffusion metric. We establish the parametrized version of Quantum Complexity and Quantum Chaos in terms of physically relevant quantities, which are not only essential in determining the stability, but also essential in providing a very significant lower bound to the generalization capability of QNN. We explicitly prove that when the system executes limit cycles or oscillations in the phase space, the generalization capability of QNN is maximized.  Finally,  we have determined the generalization capability bound on the variance of parameters of the QNN in a steady state condition using Cauchy Schwartz Inequality.
 
\vskip10pt
\hrule
\vskip10pt

\text{Keywords:~Neural~Network, ~Quantum~Complexity \& ~Chaos,~Machine~Learning.}
\vskip10pt
\vskip10pt
\end{titlepage}

\thispagestyle{empty}
\setcounter{page}{2}
\tableofcontents

\newpage
\pagenumbering{arabic}
\setcounter{page}{1}

\clearpage

\section{Introduction}

The advent of machine learning research started with the proposal of the basic framework of the neural network, perceptron by Rosenblatt in 1958. Soon in 1975, Werbos developed the back-propagation, a learning algorithm that can train multi-level perceptron. But it took almost four decades to implement a 'deep' neural network on a large industrial scale. Since then, classical machine learning has been an indispensable tool in various fields ranging from healthcare to robotics. The most crucial factor that directed the mammoth success of classical machine learning is the advancement of computational power. Back in the 1980s, Feynman proposed utilizing the quantum mechanical power of nature to compute, which led to the idea of quantum computers. But to avoid the decoherence effect in qubits, low temperature is required to sustain the coherence property of qubits. Due to these engineering issues, it took nearly three decades to build a practical quantum computer\cite{Arute2019,Chen_2018,commerce,Preskill_2018,Iverson_2020}. Recently, Google has been able to simulate quantum chemistry reactions using only 12 hydrogen atoms representing 12 qubits of information in these quantum computers \cite{Yuan1054,Arute2019}. Quantum computer researchers lately have shown interest in quantum machine learning, performing machine learning tasks using the computational power of the quantum mechanical world. There has been a lot of focus recently on the proposal of a quantum neural network as a quantum analog of a classical neural network with back-propagation as its learning algorithm \cite{Beer2020,NIPS2016_6401,PhysRevA.100.020301,Rebentrost_2019}.  

Similar to classical machine learning, quantum machine learning can be categorized into supervised, unsupervised, and semi-supervised learning like reinforcement learning. The paper focuses on the supervised learning of quantum neural networks. In a classical supervised learning algorithm, the neural network is provided with two sets of data, a training set, and a testing set. The dataset comprises ordered pairs of input and desired output, and the neural network samples these ordered pairs to the training set under a fixed probability distribution. The neural network learns from the labeled data of the training set and predicts the unlabelled data of the testing set with given accuracy and a confidence parameter. We define a loss function $f$, which represents the training error between the predicted output by the neural network and actual output in the labeled data. The neural network optimizes the loss function during the training period. In the quantum supervised learning scenario, there can be three different possible data-algorithm pairs in terms of quantum and classical nature of data and learning algorithms. In this paper, we use a hybrid quantum-classical neural network where we train quantum data with classical learning algorithms like stochastic gradient descent. In the quantum-classical hybrid framework proposed by Mitarai, a {\it Parametrized Quantum Circuits} (PQCs) is introduced by \cite{Mitarai,Benedetti_2019}. The quantum circuit is characterized by parameters that are optimized using a classical learning algorithm to optimize the loss function and simultaneously reducing the testing error. A gradient-based learning algorithm is used for back-propagation. In this paper, we have used {\it Stochastic Gradient Descent} (SGD) as our learning algorithm \cite{Bengio,chaudhari2016entropysgd,dutta2020geometry}. SGD performs gradient descent in batches rather than on the complete training set. Remarkable progress has been made in executing quantum neural networks, but there has been no significant progress in developing the quantum analogy of the statistical learning theory of neural networks. The learning theory of classical neural networks has been an important tool for computer scientists to decipher the black-box of information processing in neural networks \cite{pmlr-v97-goldfeld19a,shwartzziv2017opening,fort2019emergent,10.5555/3305381.3305487,lampinen2018analytic,chaudhari2017stochastic}. Learning theory can be analyzed from two different perspectives: statistical approach and information-theoretic approach. For classical neural networks, the work by \cite{pmlr-v97-goldfeld19a,shwartzziv2017opening} analyzed the learning dynamics using the evolution of mutual information between the output and the layers of the neural network. From a quantum analogy of the information-theoretical approach, the works by \cite{PhysRevLett.124.200504,Deutsch_2000} correlated the dynamical behavior of the tripartite information with the loss function in the training process. In this paper, we analyze from a statistical and differential geometric perspective of evaluating chaos and complexity in QNN. One of the most important aspects of learning theory is the determination of stability in the learning process of the QNN. In classical neural networks, the work \cite{lampinen2018analytic} showed the learning trajectory of deep linear networks is exponentially stable. On the other hand, there can be neural network systems with limit cycles in the phase portrait \cite{chaudhari2017stochastic}. Moreover, it is argued by \cite{YanE4185} that the coherent systems with oscillations or limit cycles are essential in the stability of continuous memory in the human brain. The stability of the learning trajectory in classical neural networks motivated us to have a look from the QNN perspective. On the other hand, over the years, scientists searched for chaos in various fields ranging from population models to black holes. The search for chaotic patterns in neuroscience has been analyzed since the inception of the Hodgkin-Huxley model \cite{doi:10.1113/jphysiol.1952.sp004764}. The work is further extended with experiments performed by the work \cite{KORN2003787}. The chaos in neural networks has been well studied over the past years \cite{Wang1990,poole2016exponential,Potapov}. The quantum processing in neurons proposed by \cite{FISHER2015593} signifies that its equivalence with the human brain, which justifies the analysis of the learning theory of quantum neural networks. Yet, there has been no significant progress in the chaos in quantum neural networks. Recently, \cite{PhysRevLett.124.200504} analyzed chaos from the notion of information scrambling in quantum neural networks. In this paper, we approach to analyze the chaos and complexity in a quantum neural network from a statistical perspective. The analysis will form a basis to further co-relate the chaos in a quantum neural network with other known physical models like black holes or economic models. This would give a deep insight into how different the human brain-motivated quantum neural networks behave from the rest of the physical models. 

A recent research direction in the classical machine learning context has been the possibility of searching for optimal neural architecture \cite{45826,luo2018neural,li2019random,pham2018efficient}. The ability to search for optimal architectures given the training dataset gives us the power to predict an optimal neural network theoretically before the long training periods. The optimality is defined on two competing terms. Firstly, the time constant of the decay of the training error, better neural network shows a larger time constant. On the other hand, one also needs to take care of the generalization error. When the neural network attains more minima in the learning manifold, the probability of reaching a wide minima increases, which would result in minimum generalization error. But for a larger time constant, the number of minima attended decreases, thus a trade-off resulting in higher generalization error. To attain this optimal ability for the neural network to reach a large number of minima with an optimal time constant, the paper attempts to map the learning trajectory or unitary evolution of QNN to the trajectory of parameters using a Riemannian manifold called diffusion metric. The diffusion metric is introduced by Foressi in \cite{Fioresi2020} and it is constructed by perturbing the flat Euclidean space by the magnitude of the noise in the gradient of the loss function. We can observe how the manifold changes in changing the neural architecture. In doing so, we will be able to correlate the optimal unitary evolution of QNN with optimal i.e. stationary action path of particles in diffusion metric. The correlation has two implications, firstly searching for an optimal QNN architecture as mentioned before, another coming from a more theoretical high energy physics perspective. The optimal unitary evolution denotes the minimum number of computations required to generate the final unitary $\mathcal{U}_f$ from the initial unitary $\mathcal{U}_i$ which in other words, the relative complexity \cite{Brown_2016,susskind2018lectures,CA} between the initial and final unitary, $\mathcal{C}(\mathcal{U}_f,\mathcal{U}_i)$. This measure of complexity in unitary space is directly correlated with the optimal trajectory evolution of parameters or geodesic in the parameter space. The correlation establishes complexity as a function of the parameters of the QNN. After establishing the complexity, an extensive study of quantum chaos has been studied \cite{Maldacena2016,Shenker2014,Swingle_2017,Swingle_2016,Swingle2018,susskind2018lectures,Brown_2018,Brown_2016}. The stability of the neural network in terms of Lyapunov stability and its evolution establishes how the neural architecture governs the stability of the neural network. Rather than an extensive study of the Lyapunov evolution, an extremal study in terms of the growth of the complexity has been carried out. 
In this connection, the out-of-time-correlator (OTOC) has also been calculated using the universality relation between complexity and OTOC, given by ${\cal C}=-\log({\rm OTOC})$~\footnote{The concept of out-of-time-correlator (OTOC) is treated as a very important probe to quantify the amplitude of quantum chaos in terms of Quantum Lyapunov exponent.  In this paper we have not explicitly computed the expression for the OTOC from the first principle,  rather using the relation between complexity and OTOC we have determined OTOC in terms of complexity for a specific region of parameter space where the relation holds good.  On the other hand,  for the other part of the parameter space where this relationship does not hold good one needs to explicitly compute the expression for the OTOC from the first principle which we have not studied in this paper. Though we have explicitly studied the behavior of the circuit complexity as a function of time for the other part of the parameter space and we have found various other non-chaotic behavior and for all of these studies, the mentioned relationship between the complexity function and OTOC is not at all valid.  This statement will be more justifiable from the performed analysis which we will show in the rest half of this paper.}

The paper considers a hybrid quantum-classical neural network framework based on PQCs \cite{Mitarai,Benedetti_2019} optimizing quantum data with classical gradient-based algorithms like stochastic gradient descent (SGD). Throughout the paper, we have assumed that the length of the training dataset is large enough for the loss function to stabilize i.e. the loss function doesn't change as we increase the length of the training set. This led us to avoid fluctuations due to sampling. The assumption of a large training dataset and the stabilizing of the loss function is inspired by Bialek \textit{et al} \cite{Bialek} corresponding to quantum computation at thermal equilibrium \cite{2018APS..MARS28001C}. The paper established the behavior of noise in SGD, which is governed by the neural architecture and dataset of QNN using the diffusion metric. The parameterized complexity of QNN is established by corresponding with the geodesic of parameter trajectories in the diffusion metric. The paper further analyses the stability of QNN using the Lyapunov exponent as a function of the neural architecture and dataset.

\begin{figure}[ht!]
\begin{center}
\includegraphics[width=17cm,height=16cm]{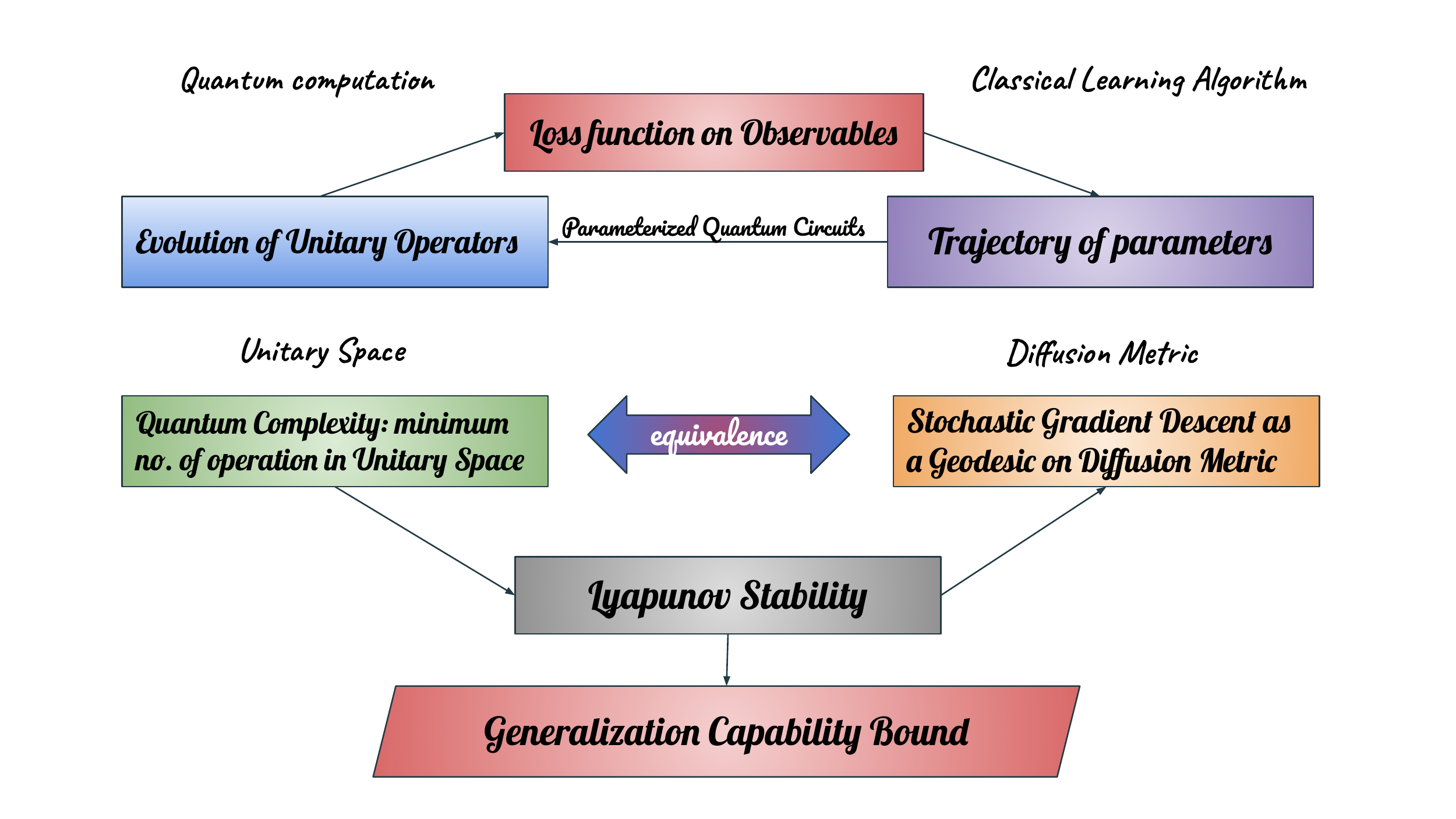}
\end{center}
\caption{Roadmap of the paper}
\end{figure}

The paper is divided into three sections, building up the mathematical background in \hyperref[sec: math background]{Section $2$} to the analysis of stability using the Lyapunov exponent in \hyperref[sec: complexity]{Section 4}. In \hyperref[sec: math background]{Section $2$}, Parameterized Quantum Circuits were introduced as a universal function approximator \cite{HORNIK1991251} and analysis was performed as a quantum analog of the statistical learning theory based on \cite{Bialek}. The diffusion metric \cite{Fioresi2020} is introduced in \hyperref[sec: metric]{Section $3$}, correlating the learning trajectory of QNN to the evolution of noise in SGD during training. After establishing the fundamental and mathematical concepts in \hyperref[sec: math background]{Mathematical Background} and \hyperref[sec: metric]{Diffusion Metric}, the paper determines complexity as a function of parameters in \hyperref[param complexity]{Parameterized Complexity}. The complexity of QNN determines the Lyapunov exponents and thus the stability analysis was established in \hyperref[sec: stability]{Quantum Lyapunov exponents}, along with \hyperref[sec: results]{Results and Discussions} showing the variations of Complexity with the architecture of the QNN.

\section{Mathematical Background }
\label{sec: math background}

We focus on executing supervised learning tasks using \textit{Parameterized Quantum Circuit} framework. In classical supervised learning, the model learns to map from an input dataset $\{x_i\}$ to the output $\{\hat y_i\}$. The map represents a $w-$parameterized function $y_i=m(x_i;w)$, which is optimized to be close to the output $\hat y_i$ for all data index $i$ belonging to the training dataset. The metric used to define the closeness of the parameterized function and output is called the loss function. The loss function can represent any metric like cross-entropy, likelihood loss, log loss, etc \cite{Bengio}. Here, we consider the mean square error loss. The main objective of the model is to optimize the loss function using a certain learning algorithm. These algorithms update the parameters $w$ of the parameterized function $m(x_i;w)$ to optimize the loss function. Here, the model optimizes the loss function employing stochastic gradient descent (SGD) as its learning algorithm. SGD is an iterative method for optimizing the loss function by using the gradient of the loss function calculated from a randomly selected subset of the training dataset \cite{chaudhari2016entropysgd,Bengio}. This supervised learning scenario is valid for both classical and quantum neural networks. In the quantum neural network context, the initial density matrix $\rho_{in}(x_i)$ is created by encoding the input data stream $\{x_i\}$ onto the \textit{Encoder Circuit} Unitary $\mathcal{U}_\phi(x_i)$, which acts on the ground state $\ket{0}$ \cite{Benedetti_2019,Mitarai}. The Unitary $\mathcal{U}_\phi$ can be represented as a sum of a linear combination of the basis operators $\alpha$ spanning over $K-$dimensional space with basis functions $\phi(x)$ as coefficients. The \textit{Encoder Circuit} Unitary $\mathcal{U}_\phi$ is characterized by basis functions $\phi_\mu(x)$, where $x$ is sampled independently and identically from the training dataset under a fixed probability distribution $P(x)$ with variance $\sigma_\eta^2$ \cite{Bialek,dutta2020geometry}. Mathematically the  unitary $\mathcal{U}_\phi$ can be represented by: 
\bea\label{eq- encoder}
\boxed{\boxed{\textcolor{BrickRed}{\textbf{\textit{Encoder Circuit Unitary : }}}\mathcal{U}_\phi(x) =\sum_{\mu=1}^K\phi_\mu(x)\sigma^\mu}}, \eea
using which the input density matrix $\rho_{in}$ is defined as:
\bea \label{eq- initial density} \boxed{\boxed{\textcolor{BrickRed}{\textbf{\textit{Input Density Matrix : }}}\rho_{\text{in}}(x_i)=\mathcal{U}_\phi(x_i)\ket{0}\bra{0}\mathcal{U}_\phi^\dagger(x_i)}}.
\eea

Here the input density matrix $\rho_{in}$ is created from the input dataset using the equation \ref{eq- initial density}. The quantum neural network (QNN) applies a parameterized Unitary operator $\mathcal{U}_{\theta}$ on the input density matrix to produce an output density matrix $\rho_{\text{out}}$ at every epoch (or iteration). Here, it is important to note that similar to the universal approximation theorem in artificial neural networks \cite{HORNIK1991251}, there always exists a quantum circuit
that can represent a target function within an arbitrarily small error. The parameterized quantum circuit learning will always be able to optimize to any arbitrary small error but the depth or complexity of the circuit increases. This optimization also doesn't guarantee the generalization capability of the quantum neural network which would result in a high testing error. Motivated by the notion of deep neural networks and the unitary arrangement proposed by Beer \textit{et. al.} \cite{Beer2020}, we use the quantum neural network architecture of stacked unitary operators with a $L$ number of layers. The unitary operator of the whole quantum circuit parameterized by $\theta$ is given by:
\bea \label{eq- unitary-theta}
\boxed{\boxed{\mathcal{U}_\theta=\prod_{i=1}^L\mathcal{U}_i}},
\eea

where $\mathcal{U}_i$ is the unitary at $i$th layer parameterized by weights $w$. The unitary at $i$th layer can be expressed as the sum of the linear combination of the basis operators $\sigma$ with $w$ as its coefficients. Mathematically, the unitary $\mathcal{U}_i$ can be expressed as follows: 
\bea \label{eq- unitary-layer}\boxed{\boxed{\mathcal{U}_i=\sum_{\nu}w^i_\nu\sigma^\nu}}.
\eea 

Combining equations \ref{eq- unitary-theta}-\ref{eq- unitary-layer}, the parameterized unitary operator can be expressed as the sum of the linear combination of the basis operators $\sigma$ spanning over $P-$dimensional space with parameter $\theta$ as its coefficient. Mathematically, the unitary operator $\mathcal{U}_\theta$ can be expressed as follows: 
\bea \label{eq- unitary}
\boxed{\boxed{\textcolor{BrickRed}{\textbf{\textit{Parameterized Unitary : }}}\mathcal{U}_\theta =\textcolor{Black}{\sum_{\nu=1}^K}\theta_\nu \sigma^\nu}},
\eea

where $\theta_\nu=g^\nu(w)$ is a function of weights. The parameters $\theta$ gets updated by the learning algorithm to optimize the loss function. The unitary operator $\mathcal{U}_\theta$ given by equation \ref{eq- unitary} acts on the initial density matrix given by equation \ref{eq- initial density} to produce the output density matrix $\rho_{\text{out}}$. We measure the output density matrix $\rho_{\text{out}}$ using the observer operator $B$ to get an expected value of the observation $B$ given by $\text{Tr}(B\rho_{out})$. QNN aims to optimize this expected observation value to a target observation value $\bar B$ as an output. For every input dataset $\{x_i\}$, we consider a corresponding output observation dataset $\{B_i\}$. During the training period, the training dataset $(x_i,\bar B_i)$ is sampled under the distribution $P(x)$. The parametrized unitary operator $\mathcal{U}_\theta$ maps the input $\{x_i\}$ encoded initial density matrix to output density matrix.
The observer maps this output density matrix to the loss function value $f$. Now using a learning algorithm (here, stochastic gradient descent) the QNN must find a sub-space of the unitary operator $\mathcal{U}_{\bar\theta}$ for which the loss function $f$ is at its minimum. But again this sub-space doesn't guarantee generalization over dataset or minimum of testing error.~\footnote{There may be an overlapping sub-sub space for which the QNN could both optimize and generalize. Note that there can also be a situation when the sub-space of optimization doesn't overlap with the sub-space of generalization. In that case, a trade-off takes place which is not favourable and a different neural architecture or \textit{Encoder Circuit} should be looked after.} The framework for Quantum Neural Network can be summarized following equations \ref{eq- encoder}-\ref{eq- initial density} and \ref{eq- unitary-theta}-\ref{eq- unitary} as: 
\bea \label{eq- rho out}
\boxed{\boxed{{\textcolor{BrickRed}{\textbf{\textit{Output Density Matrix : }}}\rho_\text{out}}(x_i)=\mathcal{U}_\theta^\dagger \rho_{\text{in}}(x_i)\mathcal{U}_\theta}}, \eea 
and \bea \label{eq- theta-f}\boxed{\boxed{\textcolor{BrickRed}{\textbf{\textit{Loss function : }}}f=\frac{1}{N}\sum_{i=1}^N\Biggl(\bar B_i-\text{Tr}(B\rho_{\text{out}}(x_i))\Biggr)^2}},
\eea

where $N$ is the length of the training dataset. We introduce a mapping $\Omega$ which the equations \ref{eq- rho out}-\ref{eq- theta-f} maps  $\Omega:\theta\mapsto f$ for every epoch. It is important to note that there is a reduction in dimension to form a surjective $K-\text{to}-1$ mapping. This results in many sub-spaces of optimal parameters for which the loss function $f$ is minimum. Under a learning algorithm (here, stochastic gradient descent) QNN changes the unitary operator every epoch $t$ by applying a mapping $\Theta:{\theta_t}\mapsto{\theta_{t+1}}$. To optimize the loss function, the learning algorithm tends to map $\Theta$ such that $\mathrm{E}(f(\theta_t))>\textcolor{Black}{\mathrm{E}(f(\theta_{t+1}))}$. SGD tends the corresponding loss function to decreases iteratively at every epoch and finally reach the unitary $\mathcal{U}_{\bar\theta}$, characterized by $\bar\theta$. In other words, when the parameters optimize $\theta\rightarrow \bar\theta$, the expected value of observation $B$ will tend towards $\bar B$, resulting in the zero training error. The parameters $\bar\theta$ are the optimal parameters. So, the matrix $\bar B$ can be represented as: 
\bea \label{eq- observation}
\boxed{\boxed{\bar B_i=\Biggl(\text{Tr}(B\mathcal{U}_{\bar\theta}^\dagger\rho_{\text{in}}^i\mathcal{U}_{\bar\theta})+\textcolor{Black}{\eta_i}\Biggr)}},
\eea

where we used a shorthand $\rho_{\text{in}}^i=\rho_{\text{in}}(x_i)$ and $\textcolor{Blue}{\eta_i}$ are the Gaussian noise with mean zero and $\sigma_\Gamma^2$ variance, similar to the classical variant used in \cite{Bialek}. The neural network optimizes by updating its parameters using  stochastic gradient descent (SGD). The learning algorithm or the mapping $\Theta:{\theta_t}\mapsto{\theta_{t+1}}$ can be given as follows:
\bea \label{eq- SGD}
\boxed{\boxed{\textcolor{BrickRed}{\textbf{\textit{Stochastic Gradient Descent : }}}\theta_{t+1}=\Biggl(\theta_t -  \frac{\Gamma}{|\mathcal{B}|}\sum_{i\in\mathcal{B}}\frac{\partial f_i}{\partial \theta},
\Biggr)}},\eea

where $\Gamma$ represents the learning rate. Rather than optimizing the whole training dataset, we optimize the dataset in batches $\mathcal{B}$, randomly picked from the whole training data. This not only reduces the computational cost but also adds stochasticity due to random sampling of batches which proves to be very essential in the generalization context, which will be discussed later in \hyperref[sec: complexity]{Complexity \& Stability}. The batch size $|\mathcal{B}|$ is much less than the total length of the training dataset $N$. The stochasticity in the SGD arises when $|\mathcal{B}|<<N$, which allows for higher stochasticity in random sampling while training. When both the length $|\mathcal{B}|\sim N$ then the stochasticity loses and SGD becomes simple (not stochastic!) gradient-descent. SGD in the last few decades has been experimentally the most efficient in terms of accuracy and computational cost. This lead to its increasing interests and documentations in the computer-science community. Recently, with the works of \cite{chaudhari2016entropysgd,Fioresi2020,dutta2020geometry}, there is a huge surge in interest in analyzing the SGD from a dynamical system perspective. We discuss this important aspect of SGD in \hyperref[sec: metric]{Diffusion Metric}. 

The loss function plays an essential part in the QNN framework as it is the objective function and purposely drives the gradient in the learning algorithm.
Combining equations \ref{eq- encoder}-\ref{eq- initial density} and equations \ref{eq- unitary}-\ref{eq- theta-f}, the loss function can be expressed as:
\bea \label{eq- f-raw}
\boxed{\boxed{f=\sigma_\eta^2+\textcolor{Black}{\sum_{\mu,\nu,\delta,\gamma}^K(\theta_\mu^*\theta_\nu-\bar\theta_\mu^*\bar\theta_\nu)(\theta_\delta^*\theta_\gamma-\bar\theta_\delta^*\bar\theta_\gamma)}\text{Tr}\Bigg(\underset{\equiv\Delta}{\underbrace{\frac{1}{N}\sum_{i=1}^N (B\sigma^\mu\rho_{\text{in}}^i\sigma^\nu\otimes B\sigma^\delta\rho_{\text{in}}^i\sigma^\gamma}})\Bigg)}}.\nonumber\\
&&
\eea

Further simplifying the factor $\Delta$ under the condition that $N\rightarrow \infty$ (large $N$) as evaluated in \cite{Bialek,dutta2020geometry} we get:
\bea \label{eq- delta}
\boxed{\boxed{\Delta=\sum_{j,k,p,q}^KA_{j'kp'q}^\infty\Biggl(B\sigma^\mu\textcolor{Black}{\sigma^j\ket{0}\bra{0}\sigma^k} \sigma^\nu\otimes B\sigma^\delta\textcolor{Black}{\sigma^p\ket{0}\bra{0}\sigma^q}\sigma^\gamma\Biggr)}}
\eea

where the expansion co-efficient $A_{j'kp'q}^\infty$ can be expressed as:\bea   \label{eq- A tensor}                   \boxed{\boxed{\textcolor{BrickRed}{\textbf{\textit{Encoder-Dataset Tensor : }}}A_{j'kp'q}^\infty=\lim_{N\rightarrow \infty}\Biggl\{\frac{1}{N}\sum_{i=1}^N \phi^{*}_j(x_i)\phi^{*}_p(x_i)\phi_{q}(x_i)\phi_{k}(x_i)\Biggr\}}}.\quad\quad\eea
The $4-$rank tensor $A^\infty$ exists under the assumption that the equation \ref{eq- A tensor} thermalizes or reaches equilibrium. The basis functions of the \textit{Encoder Circuit}  unitary operator $\mathcal{U}_\phi$ play a pivotal part in creating the initial density matrix $\rho_{\text{in}}$ in equation \ref{eq- encoder}-\ref{eq- initial density}. The basis functions are constant for a framework and thus selected prior to any training. The selection of these basis functions is thus important as the input dataset gets encoded into the basis functions. The tensor $A^\infty$ signifies the relation between the architecture of the \textit{Encoder Circuit} and dataset. Hence, we call the tensor $A^\infty$ as \textit{Encoder-Dataset} tensor. The Encoder-Dataset tensor plays the essential role to prepare the input density matrix from the input $\{x_i\}$. It may not be obvious that this tensor plays an essential role in determining the behavior of learning in QNN. But later in \hyperref[sec: stability]{Section 4.2}, we have shown that the eigenvalues of the Encoder-dataset matrix play an essential role in determining the chaotic nature of QNN. Here, we assumed that the loss function stabilizes when the number of training data is large enough to neglect the fluctuations. Continuing equation \ref{eq- f-raw} and replacing $\Delta$ using equation \ref{eq- A tensor}, \textcolor{Black}{as shown in Appendix (section \ref{eq- appendix})} the loss function takes the following simplified form:
\bea \label{eq- f}
\boxed{\boxed{
f=\sigma_\eta^2+\textcolor{Black}{\displaystyle\sum_{\mu,\nu,\delta,\gamma,j,k,p,q}^K A_{j'kp'q}^\infty(\theta_\mu^*\theta_\nu-\bar\theta_\mu^*\bar\theta_\nu)(\theta_\delta^*\theta_\gamma-\bar\theta_\delta^*\bar\theta_\gamma)}\text{Tr}\Biggl(B\sigma^\mu\textcolor{Black}{\sigma^j\ket{0}\bra{0}\sigma^k} \sigma^\nu\otimes B\sigma^\delta\textcolor{Black}{\sigma^p\ket{0}\bra{0}\sigma^q}\sigma^\gamma\Biggr)}}.\nonumber\\
\eea

The above equation shows how the loss function is governed by the selection of  \textit{Encoder-Dataset} tensor $A^\infty$, given the observation matrix $B$. An important observation from equation \ref{eq- f} is that: when the parameters optimize i.e. $\theta\rightarrow\bar\theta$, the loss function minimizes to a non-zero constant $\sigma_\eta^2$, the variance of sampling distribution $P(x)$. It is easy to mathematically validate as the loss function $f$ is a mean squared error loss, the sampling of the ordered pairs $(x_i,\bar B_i)$
 also attributed to the loss function. Consider the sampling distribution $P(x)$ as a delta function with $\sigma_\eta^2\rightarrow 0$, then the loss function $\min f$ also tends to zero. Now, when we increase the variance of the distribution, or in other words, increase the diversification of the training dataset, the $\min f$ also increases. This also shows that the neural network will fail for a uniform sampling distribution with $\sigma_\eta^2\rightarrow \infty$, there has to be an underlying structure of the dataset for the neural network to optimize.

\section{Diffusion Metric}
\label{sec: metric}

Previously in \hyperref[sec: math background]{2}, we discussed the notion of stochasticity in stochastic gradient descent (SGD) using equation \ref{eq- SGD}. The hyper-parameters of SGD i.e. the batch size $|\mathcal{B}|$ and the learning rate $\Gamma$ are crucial in achieving optimal learning trajectory in the context of computational cost and accuracy \cite{chaudhari2016entropysgd}. The learning trajectory accounts for the trajectory of parameters in the learning manifold, from the initial condition governed by the dynamical equation given by equation \ref{eq- SGD}. A faster learning rate will skip minimal in the learning manifold, thus reducing the probability of achieving better minimal for optimization and generalization. Though a smaller batch size will reduce the computational cost, it also on the other hand increase the stochastic behavior of SGD large enough to skip the minimal, increasing training and testing error. This brings the focus to find a metric to calculate the stochasticity in the SGD, to analyze the effect of the hyper-parameters, the \textit{Encoder Circuit} along the neural architecture of QNN on the behavior of stochasticity in the learning trajectory. The literature by Foressi \textit{et. al.} \cite{Fioresi2020} showed that the Diffusion matrix $D$ which is essentially the covariance matrix of the gradient of the loss function provides a great insight into the stochastic nature of SGD. The diffusion matrix $D$ becomes a null matrix when the learning trajectory is governed by a simple (not stochastic!) gradient-descent. This implies that when the matrix $D$ is null, the sampling of the batches is irrelevant to the learning algorithm. At this time, the loss function has reached its critical point or in other words, the model has learned the training dataset. The magnitude of the Diffusion matrix determines the amount of stochasticity of SGD. The work \cite{Fioresi2020} introduced a metric called Diffusion metric $\widetilde{D}$, which is created by perturbing the \textcolor{BrickRed}{Euclidean} space of parameters with the magnitude of noise in the stochastic gradient descent. Foressi \textit{et. al.} \cite{Fioresi2020} showed that the trajectory of parameters $\theta$ governed by SGD follows a geodesic path in the diffusion metric under a potential given by $V$. Mathematically, the diffusion metric is given by the following expression: 
\bea \label{eq- metric}
\boxed{\boxed{\textcolor{BrickRed}{\textbf{\textit{Diffusion Metric : }}}\widetilde{D}_{\mu\nu}=\Bigl(\delta_{\mu\nu}+\epsilon D_{\mu\nu}
\Bigr)}},\eea

where $\epsilon<1/\max\lambda_D$ where $\lambda_D$ is the set of eigenvalues of diffusion matrix $D$ and $\epsilon$ is the order of perturbation to the \textcolor{Black}{Euclidean} space in the above expression. The \textcolor{Black}{Euclidean} space corresponds to the diffusion matrix being $D=0$ or in other words, the learning trajectory is governed by simple gradient-descent. Perturbing this \textcolor{Black}{Euclidean} space with weak perturbation will distort the straight line geodesic path of parameters governed by simple gradient descent. This path corresponds to the path of the parameters with no excitation to explore other minima, thus increasing the probability of finding a better minima point resulting in better generalization. As mentioned earlier, the diffusion matrix is a covariance matrix of the gradient of the loss function $f$, which is mathematically expressed by the following expression:
\bea \label{eq- diffusion matrix}
\boxed{\boxed{\textcolor{BrickRed}{\textbf{\textit{Diffusion Matrix : }}}D_{\mu\nu}=\frac{1}{N}\sum_{i=1}^N \left(\frac{\partial f_i}{\partial \theta_\mu}\right)\left(\frac{\partial f_i}{\partial \theta_\nu}\right)-\frac{1}{N^2}\sum_{i,j=1}^N \left(\frac{\partial f_i}{\partial \theta_\mu}\right)\left(\frac{\partial f_j}{\partial \theta_\nu}\right)}}
\eea

To evaluate the diffusion matrix, we consider evaluating the gradient of the loss function $f$ as follows: 
\bea \label{eq- f-index}
\boxed{\boxed{f_i= \Big\{\textcolor{Black}{\sum_{\mu,\nu}^K(\theta_\mu^*\theta_\nu-\bar\theta_\mu^*\bar\theta_\nu)}\text{Tr}\Big(B\sigma^\mu\rho_{\text{in}}^i\sigma^\nu\Big)+\eta\Big\}^2}},\eea

using which we compute:
\bea \label{eq- grad-f}\left(\frac{\partial f_i}{\partial \theta_\zeta}\right)&=&\Big\{\textcolor{Black}{\sum_{\mu,\nu,\delta,\gamma}^K(\theta_\mu^*\theta_\nu-\bar\theta_\mu^*\bar\theta_\nu)\Big(\mathcal{\bar G}^\delta_\zeta\theta_\gamma+\mathcal{G}^\gamma_\zeta\theta_\delta^*\Big)}\text{Tr}\Big(B\sigma^\mu\rho_{\text{in}}^i\sigma^\nu \otimes B\sigma^\delta\rho_{\text{in}}^i\sigma^\gamma\Big)\nonumber\\&&~~~~~~~~~~~~~~~~~~~~~~~~~~~~~~~+2\eta\textcolor{Black}{\sum_{\delta,\gamma}^K(\mathcal{\bar G}^\delta_\zeta\theta_\gamma+\mathcal{G}^\gamma_\zeta\theta_\delta^*)}\text{Tr}\Big(B\sigma^\delta\rho_{\text{in}}^i\sigma^\gamma \Big)\Big\},\\
\left(\frac{\partial f}{\partial \theta_\zeta}\right)&=&\Big\{\textcolor{Black}{\sum_{\mu,\nu,\delta,\gamma,j,k,p,q}^K2A_{j'kp'q}^\infty(\theta_\mu^*\theta_\nu-\bar\theta_\mu^*\bar\theta_\nu)\Big(\mathcal{\bar G}^\delta_\zeta\theta_\gamma+\mathcal{G}^\gamma_\zeta\theta_\delta^*)\Big)}\nonumber\\
 &&~~~~~~~~~~~~~~~~~~~~~~~~~~~~~~~~~~~~~~~~~~~~~~~\text{Tr}\Big(B\sigma^\mu\textcolor{Black}{\sigma^j\ket{0}\bra{0}\sigma^k}\sigma^\nu \otimes B\sigma^\delta\textcolor{Black}{\sigma^p\ket{0}\bra{0}\sigma^q}\sigma^\gamma\Big)\nonumber\\&&~~~~~~~~+2\eta\textcolor{Black}{\sum_{\delta,\gamma,j,k}^KA_{j'k}^\infty\Big(\mathcal{\bar G}^\delta_\zeta\theta_\gamma+\mathcal{G}^\gamma_\zeta\theta_\delta^*\Big)}\text{Tr}\Big(B\sigma^\delta\textcolor{Black}{\sigma^j\ket{0}\bra{0}\sigma^k}\sigma^\gamma \Big)\Big\},
\eea

where the dependency of $\theta^\mu$ with respect to $\theta^\nu$ can be given by the Jacobian $\mathcal{G}^\mu_\nu$ as follows: 
\bea 
\boxed{\boxed{\textcolor{BrickRed}{\textbf{\textit{Jacobian Matrix : }}}\mathcal{G}^\mu_\nu=\left(\frac{\partial \theta^\mu}{\partial\theta^\nu}\right)=
\left\{\begin{tabular}{cc}
$\displaystyle \sum_{l\in\{L(\nu)\}} \left(\frac{\partial g^\mu(w)}{\partial w_l}\right)\left(\frac{\partial w_l}{\partial g^\nu(w)}\right)$  & $\mu\not=\nu$\\ \\
1 & $\mu=\nu$
\end{tabular}
\label{eq- jaco} 
\right\}}}\quad \quad
\eea
where $\{L(\nu)\}$ is the collection of indexes $l$ for which $\frac{\partial g^\nu(w)}{\partial w^l}\not =0$ as shown in \cite{dutta2020geometry}. It is important to note that $\mathcal{G}$ changes with epochs as weights evolve with time. The matrix $\mathcal{G}$ measures the dependence between the different parameters represented as coordinates. The matrix $\mathcal{G}$ being the Dirac-delta function infers that the parameters are independent and the parameter space corresponds to the \textcolor{Black}{Euclidean} space. From the dynamical system perspective, the matrix $\mathcal{G}$ governs the dependence of a parameter with other parameters, which in turn changes that parameter itself. One can correlate this scenario with many-body interactions with long-range hopping where the hopping energy from lattice site $i$ to $j$ corresponds to the magnitude of the matrix $\mathcal{G}^j_{i}$. When the magnitude of each element of the matrix $\mathcal{G}$ is large enough, the hopping energy is large, making the disorder strength to decrease- ergodicity arises. On the other hand, when the magnitude of each element of the matrix $\mathcal{G}$ is small enough, the hopping energy is small making the disorder strength to increase- localization arises and ergodicity is lost. In the work by \cite{chaudhari2017stochastic}, the inverse temperature $\beta$ is defined by the hyper-parameters of SGD. This motivated us to correlate the Jacobian matrix $\mathcal{G}$ with the hopping energy. The correlation provides a holistic phase diagram between temperature $T$ and disorder strength $W$ similar to any Ising-like models as shown in \cite{Altman2018,Vojta_2003}. The phase diagram will provide a deeper understanding between the equilibrium systems or non-equilibrium systems in an artificial neural network context. The study of equilibrium and non-equilibrium aspects in artificial neural networks has been discussed in \cite{chaudhari2017stochastic}. 

Similar to the assumption in the equation \ref{eq- A tensor} that the $4-$rank tensor $A^\infty$ thermalizes or reaches equilibrium, the diffusion matrix given in \ref{eq- diffusion matrix} also thermalizes. Using the similar treatment as \cite{Bialek,dutta2020geometry}, the approximated diffusion matrix can be given by 
\bea \label{eq- app diffusion matrix}
&&\textcolor{BrickRed}{\textbf{\textit{Approximated Diffusion Matrix : }}} \nonumber\\
&&D^\infty_{\zeta\eta}=\lim_{N\rightarrow\infty}D_{\zeta\eta}=4\Bigg[\textcolor{Black}{\sum_{\mu,\nu,\delta,\gamma,\omega,\kappa,\xi,\pi,j,k,p,q,r,s,a,b}^K\Big(A_{j'kp'qr'sa'b}^\infty-A_{j'kp'q}^\infty A_{r'sa'b}^\infty\Big) (\theta_\mu^*\theta_\nu-\bar\theta_\mu^*\bar\theta_\nu)}\nonumber\\&&\quad\quad\quad\quad\quad\quad\quad\quad\quad\quad\quad\quad\quad\quad\quad\quad\quad\quad\textcolor{Black}{\Big(\mathcal{\bar G}^\delta_\zeta\theta_\gamma+\mathcal{G}^\gamma_\zeta\theta_\delta^*\Big)(\theta_\omega^*\theta_\kappa-\bar\theta_\omega^*\bar\theta_\kappa)\Big(\mathcal{\bar G}^\pi_\eta\theta_\xi+\mathcal{G}^\xi_\eta\theta_\pi^*\Big)}\nonumber\\&&\quad\quad\text{Tr}\Big(B\sigma^\mu\textcolor{Black}{\sigma^j\ket{0}\bra{0}\sigma^k} \sigma^\nu\otimes B\sigma^\delta\textcolor{Black}{\sigma^p\ket{0}\bra{0}\sigma^q}\sigma^\gamma\otimes B\sigma^\omega\textcolor{Black}{\sigma^r\ket{0}\bra{0}\sigma^s} \sigma^\kappa\otimes B\sigma^\pi\textcolor{Black}{\sigma^a\ket{0}\bra{0}\sigma^b}\sigma^\xi\Big)\nonumber \\&&\quad\quad\quad+\sigma_\eta^2\textcolor{Black}{\sum_{\delta,\gamma,\pi,\xi,p,q,a,b}^KA_{p'qa'b}^\infty\Big(\mathcal{\bar G}^\delta_\zeta\theta_\gamma+\mathcal{G}^\gamma_\zeta\theta_\delta^*\Big)\Big(\mathcal{\bar G}^\pi_\eta\theta_\xi+\mathcal{G}^\xi_\eta\theta_\pi^*\Big)}\nonumber\\&&\quad\quad\quad\quad\quad\quad\quad\quad\quad\quad\quad\quad\quad\quad\quad\quad\quad\quad\text{Tr}\Big( B\sigma^\delta\textcolor{Black}{\sigma^p\ket{0}\bra{0}\sigma^q}\sigma^\gamma\otimes B\sigma^\pi\textcolor{Black}{\sigma^a\ket{0}\bra{0}\sigma^b}\sigma^\xi\Big)\Bigg], \quad\quad
\eea 

 where the matrix index $\mathcal{\bar G}^\delta_\zeta$ denotes the dependency of the complex conjugate of $\theta_\delta$ on the parameter $\theta_\zeta$. Mathematically, the matrix index is given by $\mathcal{\bar G}^\delta_\zeta=\Big(\frac{\partial\theta_\delta^*}{\partial\theta_\zeta}\Big)$. The $8-$rank \textit{Encoder-Dataset} matrix $A_{j'kp'qr'sa'b}^\infty$ can be expressed as:
 \bea                    \label{eq- A tensor 8} \boxed{\boxed{A_{j'kp'qr'sa'b}^\infty=\lim_{N\rightarrow \infty}\Biggl\{\frac{1}{N}\sum_{i=1}^N \phi^{*}_j(x_i)\phi^{*}_p(x_i)\phi^{*}_r(x_i)\phi^{*}_a(x_i)\phi_{q}(x_i)\phi_{k}(x_i)\phi_{b}(x_i)\phi_{s}(x_i)\Biggr\}}}.\eea The stochasticity of SGD changes with time which provides a temporal variation of the magnitude of perturbation in the Euclidean space. This perturbation in the approximated Diffusion metric can be correlated with the movement of masses in a Riemannian manifold where the parameters form the space-time coordinates. We now shift the problem from the approximated Diffusion metric with parameters to the trajectory of particles in the Riemannian manifold with the presence of small random masses. The magnitude of these masses is given by the magnitude of the noise in SGD, thus changes with time. The mass distribution on the Riemannian manifold also changes. Now, imagine you are told to control the trajectory of a particle from an initial point to its final point, by changing the mass distribution. The reward or aim of the particle is to visit more intermediate points while also reaching the target within a considerable time. The number of intermediate points corresponds to the generalization capability of the neural network and time here is the training time the QNN takes to reach the optimal points. In zero-mass distribution configuration, the parameters' trajectory would've been a straight line, reaching in less training time but also with less generalization capability. Changing the mass distribution increases the probability of the particle to more intermediate points, thus increasing the generalization capability. Analyzing the temporal distribution of mass thus becomes important in controlling the particle trajectory in maximizing its rewards. Realizing the correspondence, it thus becomes important to understand the stochasticity flow of SGD to control the parameter trajectory. The equation \ref{eq- app diffusion matrix} shows the dependence of neural architecture and \textit{Encoder-Dataset} tensor on the stochasticity. Thus the neural architecture and \textit{Encoder-Dataset} tensor play an important role in controlling the trajectory of parameters in the Diffusion metric, in the context of generalization capability and convergence rate. A detailed study is performed in \hyperref[sec: complexity]{Complexity \& Stability}. 
 
 Approximated diffusion matrix $D^\infty$ being a covariance matrix, is a positive semi-definite matrix with all positive eigenvalues. This property of the matrix $D^\infty$ restricts the domain of parameters $\theta$. The QNN in their whole learning trajectory should always have a parameter set $\theta$ for which $\lambda_D(\theta)\geq 0$. To visualize the limiting condition on the parameters, let us consider the parameters are equally optimized in all the directions $\textcolor{Black}{\theta_\mu^*\theta_\nu-\bar\theta_\mu^*\bar\theta_\mu=|\Delta \theta|^2}$ for all index $\mu\leq P$. The Jacobian matrix $\mathcal{G}$ and the observation matrix $B$ are considered identity matrices. The dimensions $P,K=4$ where all the basis operators are Pauli operators including the identity matrix. Using equation \ref{eq- app diffusion matrix}, the restrictions on the difference of parameters $\Delta\theta$ can be evaluated as:

 \bea \label{eq- condition diff}
D^\infty_{\zeta\eta}= &&8\text{Re}^2 \Bigg[\sum_{\mu,\nu,\omega,\kappa}^4\sum_{j,k,p,q,r,s,a,b}^4\underset{\equiv\Psi}{\underbrace{\Big(A_{j'kp'qr'sa'b}^\infty-A_{j'kp'q}^\infty A_{r'sa'b}^\infty\Big)}}\Big|\Delta \theta\Big|^2\nonumber\\&&\quad\quad\times\underset{\equiv\Phi_2}{\underbrace{\text{Tr}\Big(\sigma^\mu\textcolor{Black}{\sigma^j\ket{0}\bra{0}\sigma^k} \sigma^\nu\otimes \sigma^\zeta\textcolor{Black}{\sigma^p\ket{0}\bra{0}\sigma^q}\sigma^\zeta\otimes \sigma^\omega\textcolor{Black}{\sigma^r\ket{0}\bra{0}\alpha^s} \sigma^\kappa\otimes \sigma^\eta\textcolor{Black}{\sigma^a\ket{0}\bra{0}\sigma^b}\sigma^\eta\Big)}} \nonumber\\&&\quad\quad\quad\quad\quad\quad\quad\quad\quad\quad+\sigma_\eta^2\sum_{p,q,a,b}^4A_{p'qa'b}^\infty\underset{\equiv\Phi_1}{\underbrace{\text{Tr}\Big( \sigma^\zeta\textcolor{Black}{\sigma^p\ket{0}\bra{0}\sigma^q}\sigma^\zeta\otimes \sigma^\eta\textcolor{Black}{\sigma^a\ket{0}\bra{0}\sigma^b}\sigma^\eta\Big)}}\Bigg].
\eea

Further simplifying the factor $\Psi$ using the definitions of \textit{Encoder-Dataset} tensor in equation \ref{eq- A tensor} and \ref{eq- A tensor 8}, we get 
\bea \label{eq- cov}
\boxed{\boxed{\Psi=
\text{cov}\Big(\phi^{*}_r(x)\phi^{*}_a(x)\phi_{s}(x)\phi_{b}(x),\phi^{*}_j(x)\phi^{*}_p(x)\phi_{q}(x)\phi_{k}(x)\Big)}}
\eea

where $\text{cov}(a,b)$ is the covariance between two vectors $\vec a$ and $\vec  b$. Using the properties of Pauli matrices, the quantity $\Phi_1$ can be simplified to 
\bea \label{eq- pauli}
\Phi_1=&&4\delta_{pq}\delta_{ab}
\eea

It is important to observe than $\Phi_1$ is independent of the index $(\zeta,\eta)$. Similarly, the quantity $\Phi_2$ is also independent of the index $(\zeta,\eta)$. Thus the approximated diffusion matrix in \ref{eq- condition diff} is a constant matrix with all elements as a constant number $c$, where the quantity $c=D^\infty_{\eta\zeta}$ for all indexes $(\eta,\zeta)$ as shown in \ref{eq- condition diff}. The eigenvalues of matrix $D^\infty$ are $\lambda_D=\{0,0,0,4c\}$, which has to be a positive quantity: 

\bea \label{eq- theta condition}
0\leq &&\Bigg[\sum_{\mu,\nu,\omega,\kappa}^4\sum_{j,k,p,q,r,s,a,b}^4\text{cov}\Big(\phi^{*}_r(x)\phi^{*}_a(x)\phi_{s}(x)\phi_{b}(x),\phi^{*}_j(x)\phi^{*}_p(x)\phi_{q}(x)\phi_{k}(x)\Big)\Big|\Delta\theta\Big|^2\nonumber\\&&\text{Tr}\Big(\sigma^\mu\textcolor{Black}{\sigma^j\ket{0}\bra{0}\sigma^k} \sigma^\nu\otimes \sigma^\zeta\textcolor{Black}{\sigma^p\ket{0}\bra{0}\sigma^q}\sigma^\zeta\otimes \sigma^\omega\textcolor{Black}{\sigma^r\ket{0}\bra{0}\sigma^s} \sigma^\kappa\otimes \sigma^\eta\textcolor{Black}{\sigma^a\ket{0}\bra{0}\sigma^b}\sigma^\eta\Big)\nonumber\\&&\quad\quad\quad\quad\quad\quad\quad\quad\quad\quad\quad\quad\quad\quad\quad\quad\quad\quad\quad\quad\quad\quad\quad\quad+4\sigma_\eta^2\sum_{p,q,a,b}^4A_{p'qa'b}^\infty\delta_{pq}\delta_{ab}\Bigg].\quad\quad
\eea

The above equation may not always be true as it completely depends on the selected \textit{Encoder-Dataset} tensor $A^\infty$. Notice that when $4-$rank \textit{Encoder-Dataset} tensor $A^\infty$ has all negative values at $A^\infty_{p'pa'a}$ for all $a,p\leq 4$, then the inequality \ref{eq- theta condition} \textit{cannot} reach $|\Delta\theta|=0$. If the parameters optimize completely i.e. $|\Delta\theta|=0$ then the inequality doesn't hold true, thus a contradiction. In these cases, a stricter inequality can be evaluated according to the elemental value of the matrix $A^\infty$ and thus a limit to optimization can be evaluated. One can again correlate the inequality \ref{eq- theta condition} with particles in the Riemannian manifold where the $K=4$ parameters are the $4$ space-time coordinates. In the space-time context, the inequality \ref{eq- theta condition} shows that in certain metric (or \textit{Encoder-Dataset} tensor), the space-time coordinates can be restricted in reaching its final coordinate i.e. $\bar\theta$, making the difference $|\Delta\theta|$ a non-zero quantity. It is certainly not surprising that these types of space-time restrictions are quite common to see, reflecting an interesting correlation with QNN.

\section{Complexity \& Stability}
\label{sec: complexity}

At every epoch of training of QNN, a particular unitary operator $\cal U_\theta$ is prepared by the \textit{Parameterized Quantum Circuit}. Brown and Susskind \cite{Brown_2018,susskind2018lectures,Ag_n_2019,Stanford_2014} viewed the preparation of $\cal U_\theta$ as a time-series of discrete motions of an auxiliary particle on the Special Unitary $SU$ group space. The particle starts at the identity operator $I$ and ends at a target unitary operator $\cal U$. The complexity of the unitary operator $\cal U_\theta$ is the number of minimum operators required to create $\cal U_\theta$ by the given circuit. Mathematically, it is given by the geodesic on the $SU$ group space. QNN employs a unitary operator $\cal U_\theta$ at every epoch, thus the complexity of the QNN changes with epochs. In the QNN context, the final or target unitary operator is given by $U_{\bar\theta}$ to produce minimum training error. The particle in group space travels from the initial unitary operator $\cal U_{\theta_0}$ to the unitary operator $\cal U_{\bar\theta}$. On the other hand, this can correspond with a particle in the Diffusion metric traveling from initial parameter configuration $\theta_0$ to the optimal parameter set $\bar\theta$ as discussed in \hyperref[sec: metric]{Diffusion Metric}. A consequence of this correspondence is that the geodesic path traveled by the particle in the Diffusion metric can be correlated with the complexity as the geodesic traveled in the group space. Reflecting from the parameterized version of the approximated diffusion metric in equation \ref{eq- app diffusion matrix}, the complexity as a function of parameters can be correlated. Based on the parameterized complexity, one can further study the quantum chaos and complexity in QNN.

\subsection{Parameterized Complexity}
\label{param complexity}

The main objective of this section is to establish the complexity \cite{susskind2018lectures,Brown_2016,Stanford_2014,brand2019models,Goto_2019,Bernamonti_2020,Carmi_2017,Guo_2018,Jefferson_2017} as a function of parameters. This is motivated by the parameterized version of the approximated diffusion metric presented in equation \ref{eq- app diffusion matrix}. The parameterized complexity is evaluated by corresponding to the diffusion metric introduced in \cite{Fioresi2020}. The stochastic gradient descent follows a geodesic path on this diffusion metric, which has been discussed in \cite{Fioresi2020} is given by the following equation:
\begin{equation}
\boxed{\boxed{\textcolor{BrickRed}{\textbf{\textit{Geodesic on Diffusion Metric : }}}\left(\frac{\partial\theta}{\partial t}\right)=-(I-\epsilon D^\infty)\left(\frac{\partial f}{\partial \theta}\right)}}
\label{eq- geo}
\end{equation}
where $D^\infty$ is the approximated diffusion matrix and measures the degree of stochasticity. When the matrix $D^\infty$ becomes a null matrix, then equation \ref{eq- geo} represents the simple gradient descent as a learning algorithm.  
The optimal parameter $\bar\theta$ by integrating equation $4.1$ can be shown as: 
\bea \label{eq- optimal theta}
\boxed{\boxed{\bar\theta_\nu=\theta_0-\int_0^T \sum_{\mu=1}^4(\delta_{\mu\nu}-\epsilon D^\infty_{\mu\nu})\left(\frac{\partial f}{\partial \theta_\mu}\right) dt}}
\eea
where $T$ is a hypothetical total training time to reach the optimal parameters from the initial parameter set $\theta_0$. The equation \ref{eq- unitary} in \hyperref[sec: math background]{Mathematical Background} correlates the parameters set $\theta$ with the unitary $\cal U_\theta$. The trajectory of a particle in group space from initial unitary operator $\cal U_{\theta_0}$ exactly correspondence to the trajectory of parameters in Diffusion metric from the initial parameter set $\theta_0$ to $\bar\theta$ due to the linearity in \ref{eq- unitary}. Using this correspondence, the evolution of unitaries in the unitary space as follows:
\bea \label{eq- unitary evolve}
\boxed{\boxed{\textcolor{BrickRed}{\textbf{\textit{Evolution of Unitary : }}}\left(\frac{\partial \mathcal{U}_\theta}{\partial t}\right)
= -\sum_{\nu=1}^4\sum_{\mu=1}^4(\delta_{\mu\nu}-\epsilon D^\infty_{\mu\nu})\left(\frac{\partial f}{\partial \theta_\mu}\right)\sigma^\nu }}
\eea

Initiating with an initial unitary operator $\cal U_{\theta_0}$, the unitary $\mathcal{U}_{\theta}$ evolves with epochs tending towards the target unitary operator $\cal U_{\bar\theta}$. Susskind \cite{susskind2018lectures} discretized the special unitary group space in $\epsilon_0-$balls, where the auxiliary particle takes discretized steps into these balls to corresponds with the evolution of Unitaries. We assume a parameter set to belong to the optimal parameter set $\theta\in\bar\theta$, when the unitaries corresponding to the parameter fall in the $\epsilon_0$ balls or, in other words, $|\mathcal{U}_\theta-\mathcal{U}_{\bar\theta}|<\epsilon_0$. The work \cite{Fioresi2020} showed that SGD follows a geodesic path in diffusion metric at every epoch, which also corresponds to the complexity path on the group space. Based on the \textit{Complexity-Action conjecture} in \cite{Brown_2018,susskind2018lectures,Brown_2016,Ag_n_2019,Goto_2019,Bernamonti_2020,Carmi_2017,Jefferson_2017}, we correspond the complexity of the unitaries in the group space with the action on the diffusion metric. The \textit{Complexity-Action conjecture} as shown in \cite{Brown_2018,susskind2018lectures,Brown_2016,Ag_n_2019} is given by: 
\bea\boxed{\boxed{\textcolor{BrickRed}{\textbf{\textit{Complexity-Action Conjecture : }}}\cal C=\frac{\cal A}{\pi}}}.\eea\label{eq- CA}
The above equation shows that a change of complexity in group space will reflect a change in action in the diffusion metric and vice versa. The definition of Action defined on the diffusion metric \cite{Fioresi2020} is given by \bea \label{eq- action}\boxed{\boxed{\textcolor{BrickRed}{\textbf{\textit{Action : }}}\mathcal{A}=\int \Biggl[\sqrt{\sum_{\mu,\nu}\widetilde{D}^\infty_{\mu\nu}\dot\theta^\mu\dot\theta^\nu}-V\Biggr]dt\hspace{15mm}\text{where,}\hspace{2mm}V=-\int_{\theta}^{\bar\theta}\sum_\mu\Biggl[\frac{\partial}{\partial t}\Biggl(\frac{\partial f}{\partial \theta}\Biggr)\Biggr]d\theta}}\nonumber\\
&&\eea  
From the \textit{Parameterized Quantum Circuit} perspective, a change in the complexity will ensure a change in the unitary operator. This change in the unitary operator will cause a change in the parameter configuration in the diffusion metric, thus changing the action in the metric. Using equations \ref{eq- unitary evolve} and \ref{eq- action}, the unitary as a function of complexity can be written as:
\bea \label{eq- uc 1}
\boxed{\boxed{\left(\frac{\partial\mathcal{U}}{\partial\mathcal{C}}\right)=\frac{1}{\pi\Biggl[\displaystyle \sqrt{\sum_{\mu,\nu}\widetilde{D}^\infty_{\mu\nu}\dot\theta^\mu\dot\theta^\nu}-V\Biggr]}\sum_{\nu,\mu=1}^4\Big(\delta_{\mu\nu}-\epsilon D^\infty_{\mu\nu}\Big)\left(\frac{\partial f}{\partial \theta_\mu}\right)\sigma^\nu}}. \quad\quad\quad
\eea
where $V$ is the potential under which the parameters evolve. On the other hand, using equation \ref{eq- unitary} the gradient of unitary with respect to change in complexity is given by:
\bea \label{eq- uc 2}
\boxed{\boxed{\left(\frac{\partial\mathcal{U}}{\partial\mathcal{C}}\right)
=\sum_{\mu}^4\sum_\nu^4\frac{\mathcal{G}^\nu_\mu\sigma^\nu}{\mathcal{C}_\theta^\mu}}}
\eea
where  we have introduced the following quantity:
\bea\mathcal{C}^\mu_\theta=\left(\frac{\partial \mathcal{C}}{\partial\theta_\mu}\right)\eea \label{eq- complexity element}
 Therefore, equating the equations \ref{eq- uc 1} and \ref{eq- uc 2}, one can conclude:
\bea \label{eq- param complexity 1}
\boxed{\boxed{\left(\frac{\partial \mathcal{C}}{\partial \theta_\mu}\right)=\frac{\pi\Biggl[\displaystyle\sqrt{\sum_{\zeta,\eta}\widetilde{D}^\infty_{\zeta\eta}\dot\theta^\zeta\dot\theta^\eta}-V\Biggr]}{\displaystyle\sum_\nu\Big[\mathcal{G}^{-1}(I-\epsilon D^\infty)\Big]_{\mu\nu}\left(\frac{\partial f}{\partial\theta_\nu}\right)}}} 
\eea
The above equation establishes the distribution of complexity as a function of parameter and $\Big[\mathcal{G}^{-1}(I-\epsilon D^\infty)\Big]_{\mu\nu}$ represents the elemental value of the matrix $\mathcal{G}^{-1}(I-\epsilon D^\infty)$. The complexity is thus given by the following expression:
\bea \label{eq- param complexity 2}
\boxed{\boxed{\textcolor{BrickRed}{\textbf{\textit{Parameterized Complexity : }}}\mathcal{C}(\theta)=\pi\sum_\mu \int^{\theta}_{\theta_0} \Biggl[\frac{\sqrt{\sum_{\zeta,\eta}\widetilde{D}^\infty_{\zeta\eta}\dot\theta^\zeta\dot\theta^\eta}-V}{\displaystyle\sum_\nu\Big[\mathcal{G}^{-1}(I-\epsilon D^\infty)\Big]_{\mu\nu}\left(\frac{\partial f}{\partial\theta_\nu}\right)} \Biggr]d\theta_\mu}}\nonumber\\
\eea
 
 The above expression of Complexity $\mathcal{C}$ is difficult to evaluate exactly, analytically. We evaluate the complexity at specific epochs of the learning trajectory, as established in the next sub-section \ref{sec: stability}. 
 
\subsection{Stability analysis using Quantum Lyapunov exponents}
\label{sec: stability}

\begin{figure}[ht!]
\begin{center}
\includegraphics[width=17cm,height=12cm]{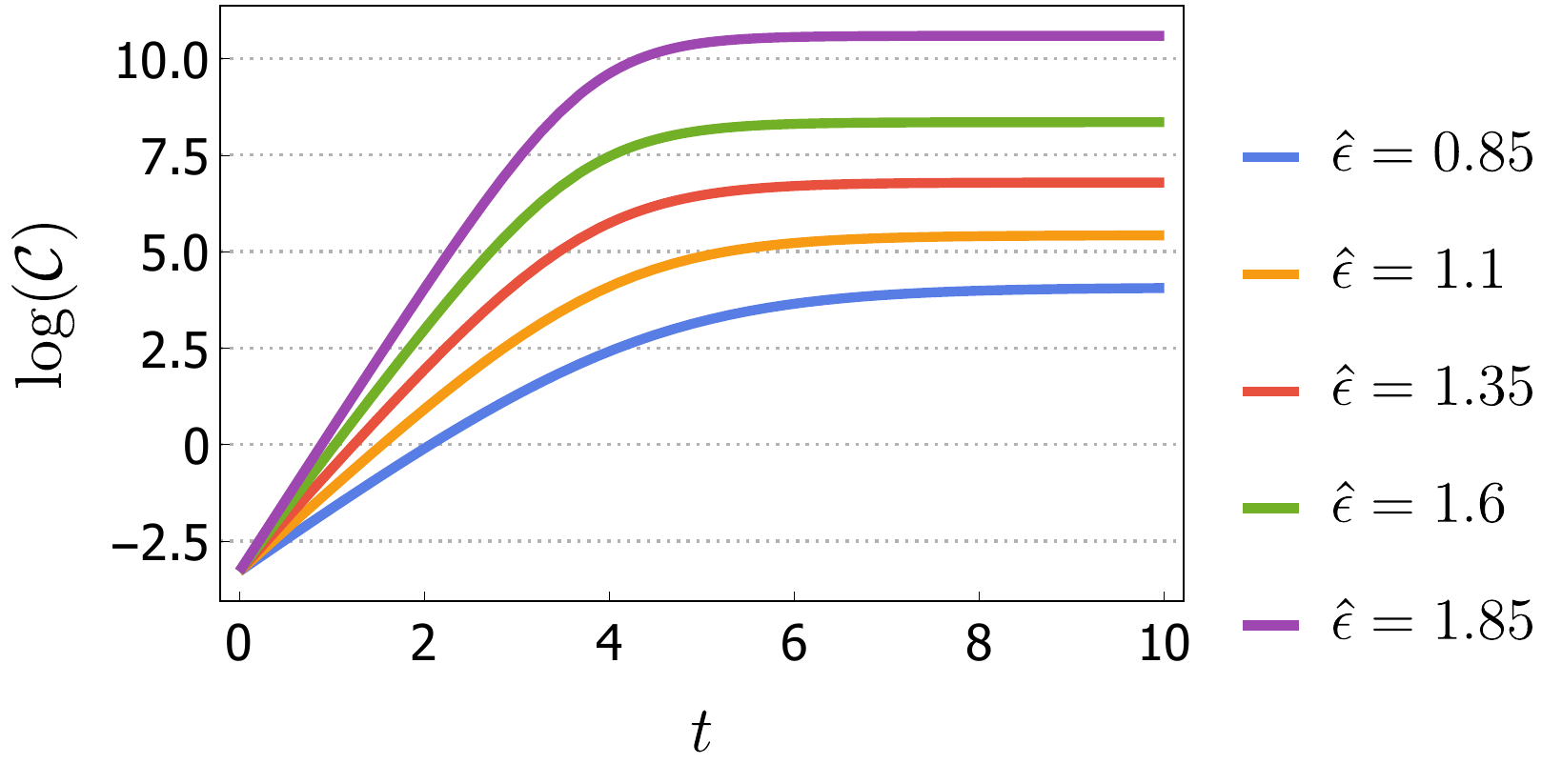}
\end{center}
\caption{Evolution of logarithm of Complexity for a $K=2$ parameter system with a variation of eigenvalue of \textit{Encoder-Dataset} tensor $A^\infty$}
\label{fig- logC}
\end{figure}

\begin{figure}[ht!]
\begin{center}
\includegraphics[width=17cm,height=8cm]{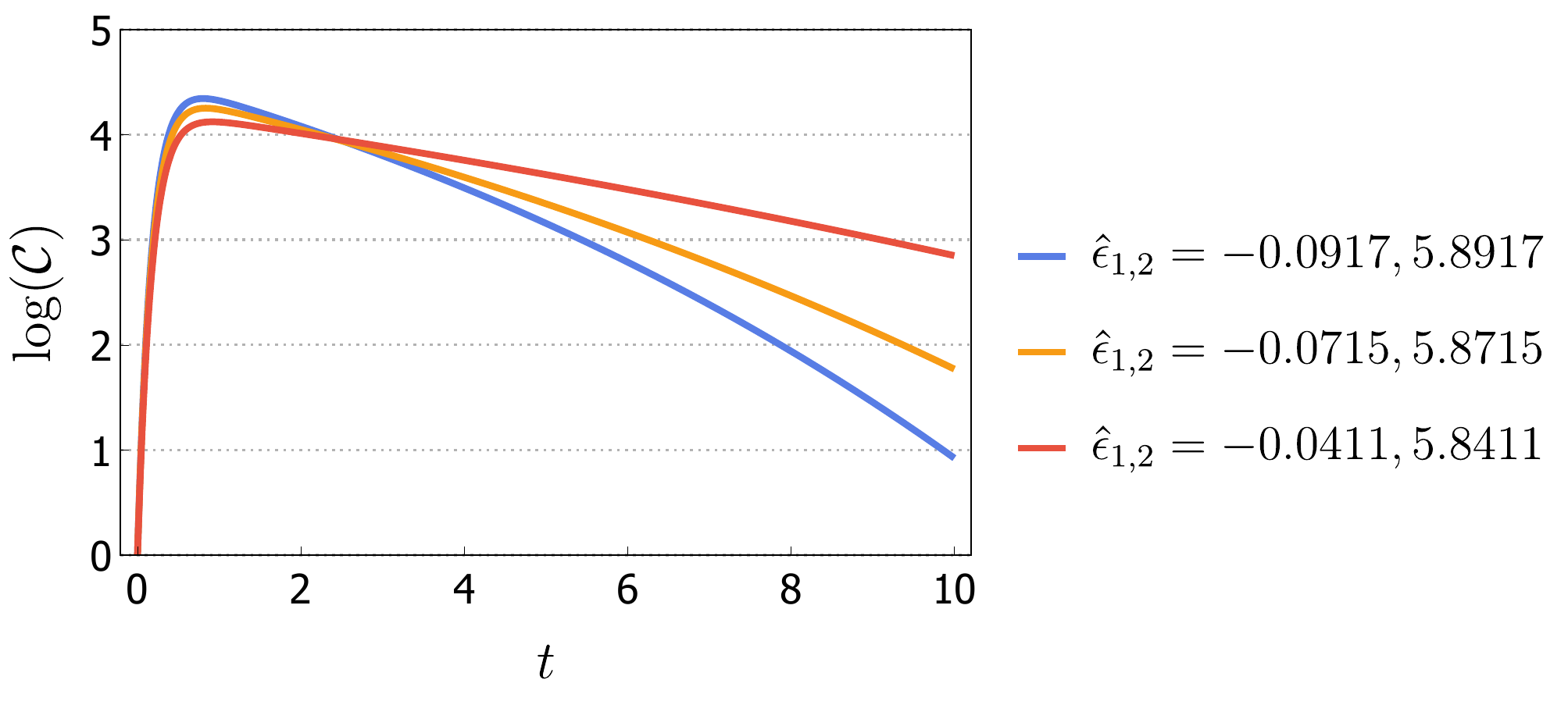}
\end{center}
\caption{Evolution of Complexity for a $K=2$ parameter system with a variation of eigenvalue of \textit{Encoder-Dataset} tensor $A^\infty$ in a non-chaotic regime}
\label{fig- complexity_dec}
\end{figure}

\begin{figure}[ht!]
\begin{center}
\includegraphics[width=17cm,height=8cm]{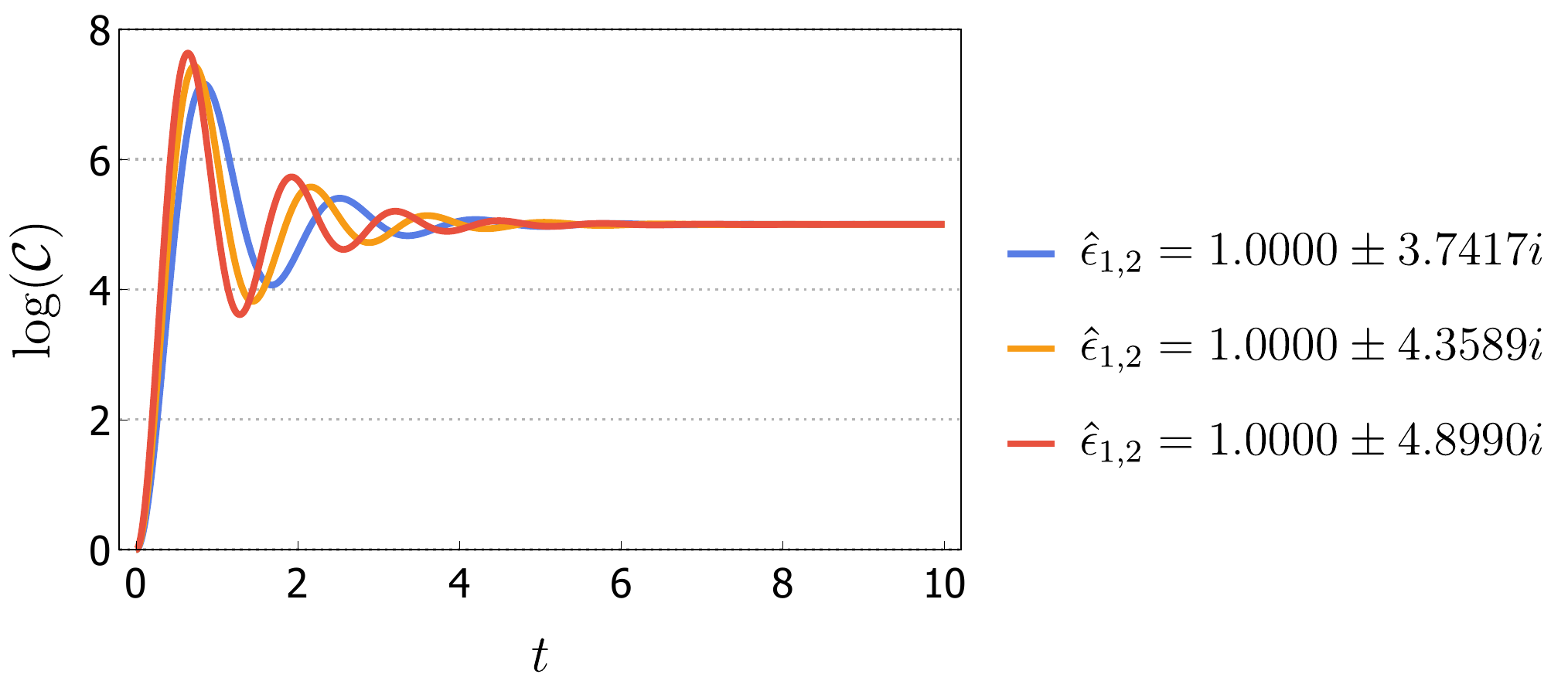}
\end{center}
\caption{Evolution of Complexity for a $K=2$ parameter system with a variation of eigenvalue of \textit{Encoder-Dataset} tensor $A^\infty$ in a non-chaotic regime. The non-zero complex part of the eigenvalues of $A^\infty$ gives rise to the oscillatory behavior of the Complexity.}
\label{fig- complexity_osc}
\end{figure}

\begin{figure}[ht!]
\begin{center}
\includegraphics[width=17cm,height=8cm]{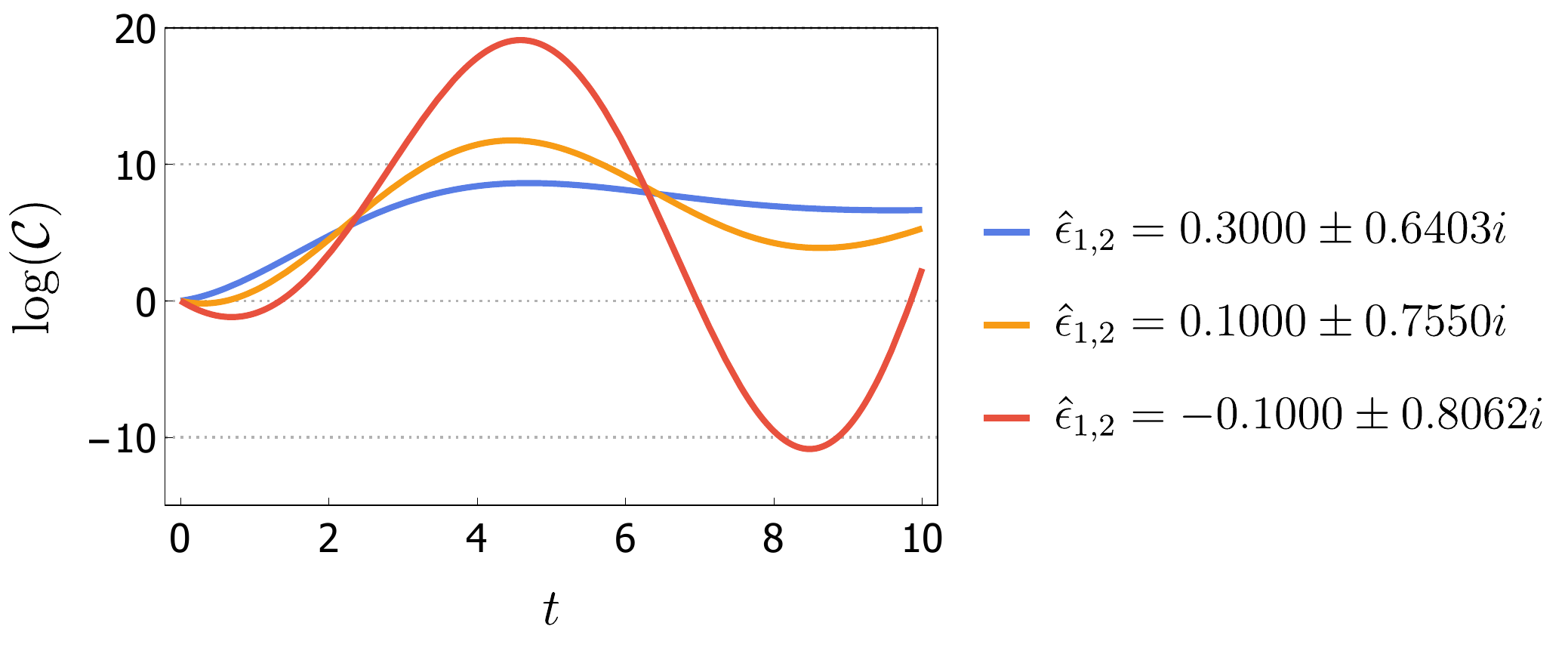}
\end{center}
\caption{Evolution of Complexity for a $K=2$ parameter system with a variation of eigenvalue of \textit{Encoder-Dataset} tensor $A^\infty$ in a non-chaotic regime. The non-zero complex part of the eigenvalues of $A^\infty$ gives rise to the oscillatory behavior of the Complexity.}
\label{fig- complexity_osc_2}
\end{figure}

\begin{figure}[ht!]
\begin{center}
\includegraphics[width=17cm,height=8cm]{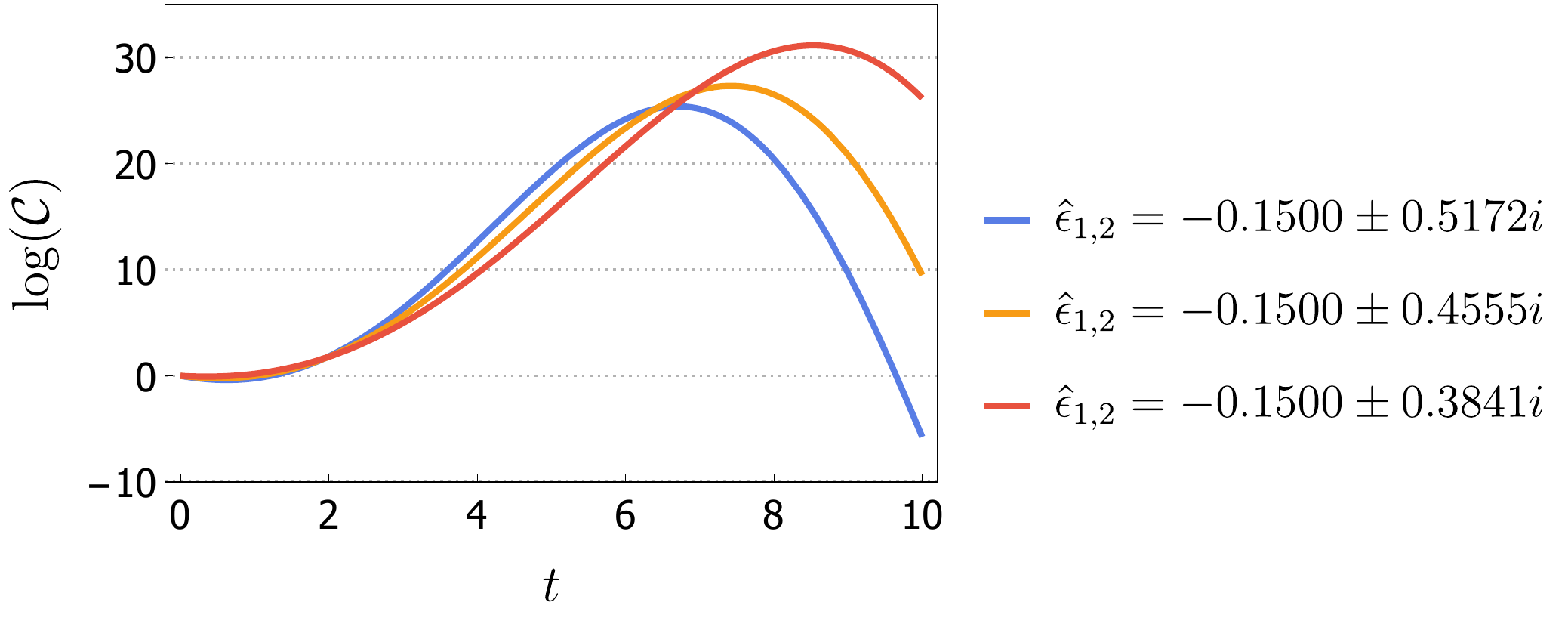}
\end{center}
\caption{Evolution of Complexity for a $K=2$ parameter system with a variation of eigenvalue of \textit{Encoder-Dataset} tensor $A^\infty$ in a non-chaotic regime. The non-zero complex part of the eigenvalues of $A^\infty$ gives rise to the oscillatory behavior of the Complexity.}
\label{fig- complexity_osc_3}
\end{figure}

\begin{figure}[ht!]
\begin{center}
\includegraphics[width=17cm,height=8cm]{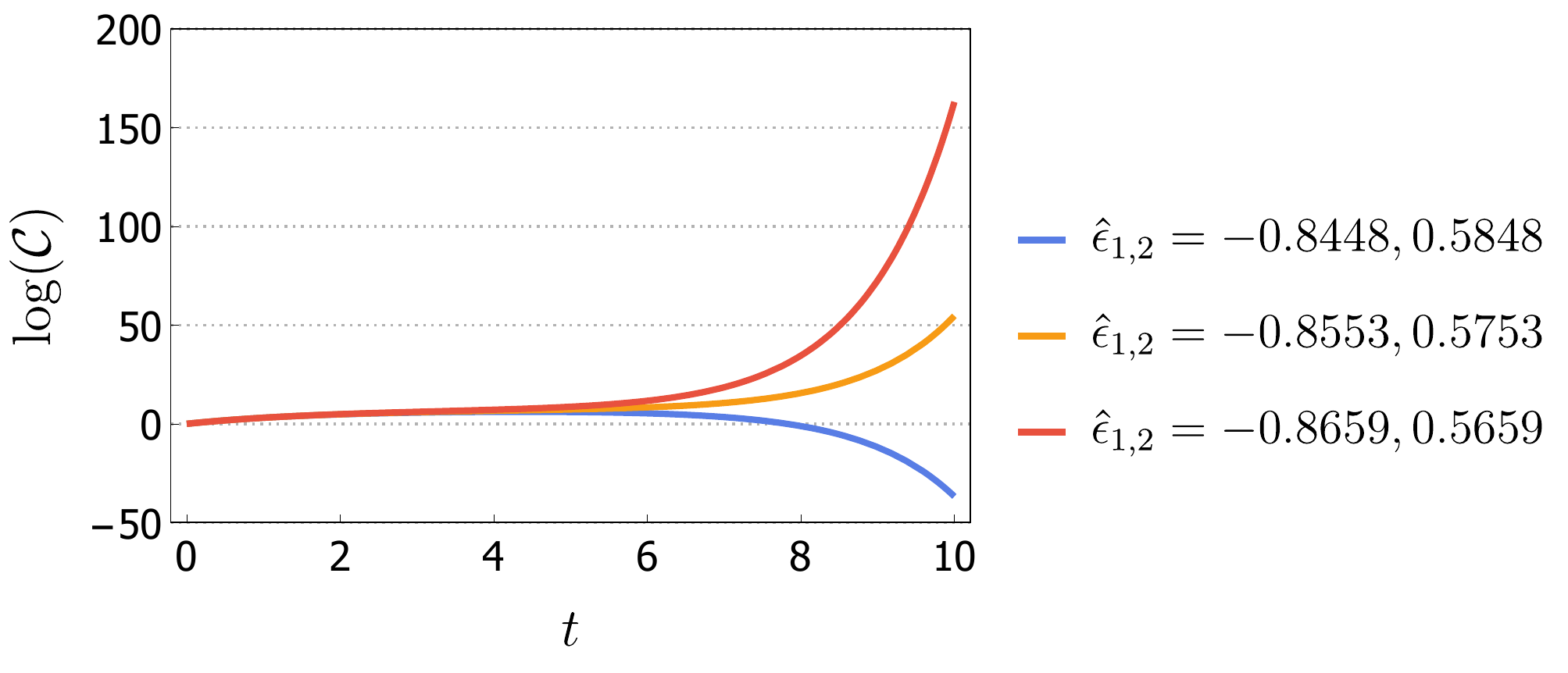}
\end{center}
\caption{Evolution of Complexity for a $K=2$ parameter system with a variation of eigenvalue of \textit{Encoder-Dataset} tensor $A^\infty$ in a non-chaotic regime}
\label{fig- complexity_dec_2}
\end{figure}

Using \cite{Bhagat:2020pcd,Swingle_2016,Swingle2018,Choudhury_2020,2017APS..DMP.T7009B,Bhargava:2020fhl}, one can write down the following relation between OTOC and complexity:
\bea \label{eq- otoc}\boxed{\boxed{{\cal C}=-\log({\rm OTOC})\quad \quad {\rm and}\quad {\rm OTOC}=\exp\left(-\exp(\lambda \theta)\right)}},\eea 
The computed Complexity function and the OTOC are connected through this consistency relation as shown in equation \ref{eq- otoc}. But it is important to note that the relationship holds for the parameter space where the exponential growth-like feature can be observed for the time scale. This is not a universal relation. In general, there will be parameter sets for which this desired exponential growth and then saturation bound-like behavior cannot be observed. In these cases, to compute OTOC, one needs to explicitly compute OTOC as stated previously \cite{Bhagat:2020pcd,Bhargava:2020fhl} and not by connecting complexity and OTOC as shown in equation \ref{eq- otoc}. Without doing an explicit rigorous computation of the OTOC function in the present context, we want to extract the information regarding the quantum 
chaotic phenomena from the obtained numerical time-dependent solution of the
Complexity function. For other parameter values involved in the complexity of
the function we cannot able to get such a desired chaotic feature and consequently
in that regime, it is not at all possible to connect the Complexity function with
the OTOC function just by using a simple relationship. Additionally, it is
important to note that, in this desired chaotic regime, for the given choice of
the parameter values, the analyticity of the OTOC allows us to numerically
implement the connection between the Chaotic function and OTOC function
using the above mentioned relation \ref{eq- otoc}. For other ranges of the parameter
values, this is not at all true. We have computed complexity for which the Complexity $\mathcal{C}-\text{OTOC}$ relationship holds namely, Figure \ref{fig- logC}. But on the other hand, we also gave examples for which exponential growth is not observed namely, Figures \ref{fig- complexity_dec}-\ref{fig- complexity_dec_2}, for which the relation \ref{eq- otoc} doesn't hold. The equation \ref{eq- otoc} further implies the following combined universal relation which particularly holds good in the context of the quantum description of chaotic phenomena:
\bea \label{eq- lyapunov}\boxed{\boxed{\textcolor{BrickRed}{\textbf{\textit{Complexity : }}}{\cal C}=-\log({\rm OTOC})=\exp(\lambda \theta)}},\eea
where $\lambda$ is identified to be the Lyapunov exponent.

Similar to the argument made above,  this relation is only valid for the parameter set for which there is exponential growth. There may be other parameter space for which complexity may have a different later time behavior like oscillatory (Figures \ref{fig- complexity_osc}-\ref{fig- complexity_osc_3}) nature.  For cases when there is exponential growth, one can compute
the Lyapunov exponent from the Complexity plot by using:  
\bea \label{eq- MSS 1}\boxed{\boxed{\textcolor{BrickRed}{\textbf{\textit{Lyapunov Exponent : }}}\lambda=\left(\frac{\partial \log(\mathcal{C})}{\partial \theta}\right)}}
\eea 
which is basically can be measured from the slope of the $\log(\mathcal{C})$ vs $\theta$ plot, which we have plotted in the later half of this paper. 


In equation \ref{eq- param complexity 2}, the complexity expression is difficult to evaluate. We analyze complexity in certain critical learning epochs, when the system is in a steady state i.e. the velocity of parameters $\dot\theta=0$. Note that the set of parameters when the loss function is at a local minimum, the system is necessarily in a steady state using equation \ref{eq- geo}. We introduce a steady-state parameter set $\theta_{ss}$ during which the system is stationary. So the optimal parameter $\bar \theta$ belongs to the parameter set $\theta_{ss}$. ~\footnote{If the loss function $f$ has only one minimum i.e. its global minimum, then the steady state $\theta_{ss}$ set will only contains the optimal parameter $\bar\theta$.} Thus, in the framework of QNN the Lyapunov exponent is evaluated from the complexity for these epochs by the following simplified expression:
\bea \label{eq- lyap}
\boxed{\boxed{\vec\lambda=\frac{1}{\displaystyle\int_{\theta_0}^{\theta_{ss}} \sum_\mu \frac{d\theta_\mu}{\displaystyle\sum_\nu\Big[\mathcal{G}^{-1}(I-\epsilon D^\infty)\Big]_{\mu\nu}\left(\frac{\partial f}{\partial\theta_\nu}\right)}  }\times\Bigg[\frac{1}{\displaystyle\sum_\nu\Big[\mathcal{G}^{-1}(I-\epsilon D^\infty)\Big]_{\mu\nu}\left(\frac{\partial f}{\partial\theta_\nu}\right)\Bigg|_{\theta=\theta_{ss}}}\Bigg]_{\vec\mu}}},\nonumber\\
&&
\eea
where $[.]_{\vec\mu}$ represents vector running through the parameter $\mu$. Using equation \ref{eq- lyap}, it is evident that $\text{sign}\Big(\displaystyle\sum_\nu\Big[\mathcal{G}^{-1}(I-\epsilon D^\infty)\Big]_{\mu\nu}\left(\frac{\partial f}{\partial\theta_\nu}\right)\Bigg|_{\theta=\theta_{ss}}\Big)$ determines the nature of the system. It is important to note that there can be a case of inflection in terms of the system stability when the Lyapunov exponent changes its sign. The Lyapunov exponent takes an intermediate form when the matrix $\displaystyle\mathcal{G}^{-1}(I-\epsilon D^\infty)\left(\frac{\partial f}{\partial\theta}\right)\Bigg|_{\theta=\theta_{ss}}$ becomes singular. For the further computational simplification purpose we introduce a function, $p(\theta)$ which is defined as:
\bea \label{eq- p}
\boxed{\boxed{ p(\theta):\equiv\mathcal{G}^{-1}(I-\epsilon D^\infty)\left(\frac{\partial f}{\partial\theta}\right)}}
\eea
We analyse the equation \ref{eq- lyap} by tending the quantity $p$ at its extremal values i.e. initially we analyse when $p\rightarrow 0$ and then we analyse when $p\rightarrow \infty$. Using this mentioned identification we get the following simplified expression for the Lyapunov exponent:
\bea \label{eq- lyap reduced}
\boxed{\boxed{\lambda=\left(\frac{1}{\displaystyle p(\theta_{ss})\int_{\theta_0}^{\theta_{ss}}\frac{d\theta}{p(\theta)}}\right)=\left(\frac{1}{\displaystyle  \theta_{ss}-\theta_0+\int_{\theta_0}^{\theta_{ss}}\underset{\equiv {\cal K}(\theta)}{\underbrace{\Big(p'(\theta)\int\frac{d\theta}{p(\theta)}\Big)}}d\theta}\right)}},
\eea
where we introduce a new function ${\cal K}(\theta)$, which is given by the following expression:
\bea \label{eq- k}\boxed{\boxed{{\cal K}(\theta)=p'(\theta)\int\frac{d\theta}{p(\theta)}}}.\eea
During this simplification, we have used the integration by parts in the above mentioned second step.

This paper considers two extreme situations i.e. when the quantity $p(\theta)\rightarrow 0$ and $p(\theta)\rightarrow\infty$.
Initially, a conditioned analysis on the stability of the system is performed here using $p(\theta)\rightarrow 0$:
\bea \label{eq- lyap 0}
\boxed{\boxed{\lambda=\lim_{p\rightarrow 0}\left(\frac{1}{\displaystyle p(\theta_{ss})\int_{\theta_0}^{\theta_{ss}}\frac{d\theta}{p(\theta)}}\right)=\lim_{p\rightarrow 0}\left(\frac{1}{\displaystyle  \theta_{ss}-\theta_0+\int_{\theta_0}^{\theta_{ss}}\underset{\equiv {\cal K}(\theta)}{\underbrace{\Big(p'(\theta)\int\frac{d\theta}{p(\theta)}\Big)}}d\theta}\right)}},
\eea
Using the equation, the following observations can be made when $p\rightarrow 0$:
\begin{enumerate}
\item If $p'(\theta)$ is a negative finite quantity at the critical parameter set $\theta^*$ when $p(\theta^*)=0$, then Lyapunov exponent tends towards $0^-$, thus the system stabilises with oscillations or limit cyclic behavior in the phase space. 
\item While if $p'(\theta)$ is a positive finite quantity at the critical parameter set $\theta^*$ when $p(\theta^*)=0$, then Lyapunov exponent tends towards $0^+$, where the chaotic nature of the system arises with unstable limit cycles. 
\item If $p'(\theta)=0$, then ${\cal K}(\theta)$ can be further simplified to be following form:  
\bea \label{eq- k simple}
\boxed{\boxed{{\cal K}(\theta)=\log(p(\theta))+\int \Big(p''(\theta)\int\frac{d\theta}{p(\theta)}\Big)d\theta}}.
\eea
When $p''(\theta)=0$, the value of ${\cal K}(\theta)\rightarrow -\infty$ thus the Lyapunov exponent $\lambda\rightarrow 0^-$, stabilizing the system with limit cycles. 
\end{enumerate}
So, the condition with the quantity $p(\theta)\rightarrow 0$, shows the system inherently can execute stable or unstable limit cycles in the phase space. 

Now, let us consider another limiting condition with $p(\theta)\rightarrow \infty$, the Lyapunov exponent is given by the following simplified expression:
\bea \label{eq- lyap infinity}
\boxed{\boxed{\lambda=\lim_{p\rightarrow \infty}\left(\frac{1}{ \displaystyle p(\theta_{ss})\int_{\theta_0}^{\theta_{ss}}\frac{d\theta}{p(\theta)}}
\right)=\lim_{p\rightarrow \infty}\left(\frac{1}{ \displaystyle\theta_{ss}-\theta_0+\int_{\theta_0}^{\theta_{ss}}\underset{\equiv {\cal K}(\theta)}{\underbrace{\Big(p'(\theta)\int\frac{d\theta}{p(\theta)}\Big)}}d\theta}\right)\approx\frac{1}{\theta_{ss}-\theta_0}}}\nonumber\\
&&
\eea
which can be further generalized to the set of Quantum Lyapunov exponents $\displaystyle \vec\lambda=\Big[\frac{1}{\theta_{ss}-\theta_0}\Big]_{\vec\mu}$ from the obtained result. Now, here it is important to note that the result here we have obtained is actually irrespective of $p'(\theta)$. 
Using the equation \ref{eq- lyap infinity}, it is evident that when $\theta\rightarrow \bar\theta$, the Lyapunov exponent $\lambda\rightarrow\infty$. 

An optimization limit can be achieved from equation \ref{eq- lyap infinity},  by analyzing the different learning epochs of the training phase.  Optimization for the artificial neural network(ANN) with a high number of trainable parameters is empirically easy and can fit any training dataset resulting in zero training data. But the zero training error doesn't assure the generalization capability of neural networks  \cite{li2020understanding,zhang2016understanding,10.5555/3295222.3295344}. When a neural network generalizes over a dataset, it understands the underlying structure of the dataset and thus reduces the difference between training error and testing error, called generalization error. Not much is discussed in the context of generalization or optimization in QNN as compared to ANN. Recently, \cite{jiang2020generalization} showed that QNN with the same structure as the corresponding ANN will have a better generalization property. In this paper, we mainly discussed or focused on the optimization property of QNN. We focused on the trajectory of parameters to their optimal parameter set in the learning manifold. But this is focusing on one-half of the portion training error. We introduce the generalization capability of QNN inspired by the notion of generalization in ANN as discussed in \cite{jiang2020generalization,li2020understanding,zhang2016understanding,chaudhari2016entropysgd,shwartzziv2017opening,fort2019emergent}. As the QNN framework used in the paper is a quantum-classical hybrid, where the parameters are optimized in classical SGD, we can use the concepts of generalization in ANN. Thereby, we focused on the fact that the generalization capability of a neural network is associated with the variance of the parameters \cite{chaudhari2016entropysgd,chaudhari2017stochastic}, higher generalization capability has high variance. Intuitively, it is taking into account the fact that a higher variance of parameters will increase the probability of finding a better optimal point which would result in better generalization capability. Thus the variance of parameters is a measure of the generalization in neural networks. The variance of parameters when the system is in a steady-state condition is given by the following expression:
\bea
\sigma_\theta^2\Big|_{\theta=\theta_{ss}} &=&\frac{1}{K}\sum_\mu^K (\theta_\mu-\widehat{\theta})^2\Big|_{\theta=\theta_{ss}}\nonumber\\
&\geq & \frac{K}{\displaystyle \left(\sum_\mu^K \frac{1}{(\theta_\mu-\widehat{\theta_0})^2}\right)\Bigg|_{\theta=\theta_{ss}}}+|\widetilde{\delta\theta}|^2+\frac{2K|\widetilde{\delta\theta}|}{\displaystyle \left(\sum_\mu^K \frac{1}{(\theta_\mu-\widehat{\theta_0})}\right)\Bigg|_{\theta=\theta_{ss}}} \nonumber\\
&& \eea
where $\widetilde{\delta\theta}=\theta_{ss}-\theta_0$ and in the last step we have used the well known {\it Cauchy Schwartz Inequality}.  After working out a bit we derive the following bound on the variance:
\bea\label{eq- generalization} \boxed{\boxed{\textcolor{BrickRed}{\textbf{\textit{Generalization Capability Bound : }}}\sigma_\theta^2\Big|_{\theta=\theta_{ss}}\geq  \frac{K}{\text{Tr}(\lambda\lambda^T)}+\frac{1}{\lambda^T\lambda}+\frac{2K}{\sqrt{\lambda^T\lambda}\text{Tr}(\sqrt{\lambda\lambda^T})}}}.\nonumber\\
&&
\eea
where $\widehat{\theta}$ is the averaged parameters over its indexes.  The inequality \ref{eq- generalization} shows that when the Lyapunov exponent is minimum or $\lambda\rightarrow 0$, then the generalization capability is at its maximum. This is shown as when the system shows limit cycles with $\lambda\rightarrow 0$ in phase space, the generalization capability reaches maximum. Interestingly, the work by \cite{YanE4185} also argued that these oscillations in phase space are a crucial part of the stability of continuous memories in the human brain. The inequality \ref{eq- generalization} gives a theoretical perspective to this argument. Moreover, inequality \ref{eq- generalization} also shows that with an increase in inverse temperature $\beta$, the generalization capability increases. Correlating with the phase diagram \cite{Altman2018,Vojta_2003}, this corresponds to many-body localization or non-equilibrium states as the work \cite{chaudhari2017stochastic} showed for an artificial neural network. Along with oscillations, \cite{YanE4185} argued coherent phases like many-body localization also play a pivotal role in the stability of continuous memories. But on the other hand, increasing the inverse temperature also corresponds to a slower convergence rate as shown in the equation \ref{eq- SGD}. Thus there is a trade-off between convergence rate and generalization capability as previously intuitively mentioned.

The change in the nature of the Lyapunov exponent is due to the matrix $\mathcal{G}^{-1}$. For $p\rightarrow 0$, the matrix $\mathcal{G}\rightarrow \infty$ and the system can be unstable or stable limit cycles depending on gradient $p'$ with no significant chaos with maximum generalization capability. But for $p\rightarrow\infty$, using the Lyapunov exponent, the complexity $\mathcal{C}$ \cite{Bhagat:2020pcd} can be given as:
\bea \label{eq- complexity reduced}
\boxed{\boxed{\frac{\partial \log(\mathcal{C})}{\partial \theta} =\frac{1}{\theta_{ss}-\theta_0}
\quad\quad\Longrightarrow\quad \quad\mathcal{C}= k\prod_\mu\exp\Big(\frac{\theta_\mu}{\theta_{ss\mu}-\theta_{0\mu}}\Big)}}
\eea
where $\theta_{ss,\mu}$ is the $\mu-$index of the steady state parameter $\theta_{ss}$, which in this case is taken to be  and $k$ is the constant to integration.

The out-of-order correlator $\text{OTOC}$ \cite{Swingle_2016,Swingle2018,Choudhury_2020,2017APS..DMP.T7009B,Bhargava:2020fhl,Maldacena2016,Choudhury_2021c,BenTov:2021jsf} is given by $\text{OTOC}=\exp(-\mathcal{C})$ which shows that out-of-order correlator can be represented as: 
\begin{equation}
\boxed{\boxed{\text{OTOC}=\exp\Bigg[-k\prod_\mu\exp\Big(\frac{\theta_\mu}{\theta_{ss,\mu}-\theta_{0,\mu}}\Big)\Bigg]}}.
\label{eq- otoc eva}
\end{equation}
The scrambling time $t_*$ as shown in \cite{Swingle_2017,Swingle_2016,Swingle2018,Lashkari_2013,susskind2018lectures,Bhargava:2020fhl} is given by: 
\begin{equation}
\boxed{\boxed{t_*=\Big(\theta_{ss}-\theta_0\Big)\log(N)}}.
\label{eq- scramb}
\end{equation}
The entropy $S$ as shown in \cite{Bhagat:2020pcd,Bhargava:2020fhl,susskind2018lectures,Brown_2018} can be expressed as:
\begin{equation}
\boxed{\boxed{S=\beta\left(\frac{\partial \mathcal{C}}{\partial t}\right)=\frac{2k|\mathcal{B}|}{\Gamma}\sum_\mu\Big(\frac{\dot\theta_\mu}{\theta_{ss,\mu}-\theta_{0,\mu}}\Big)\exp\Big(\frac{\theta_\mu}{\theta_{ss,\mu}-\theta_{0,\mu}}\Big)\prod_{\nu\not=\mu}\exp\Big(\frac{\theta_\nu}{\theta_{ss,\nu}-\theta_{0,\nu}}\Big)}}.
\label{eq- entropy eva}
\end{equation}

where $\beta$ is the inverse temperature as expressed previously in the classical machining learning context \cite{chaudhari2016entropysgd},\cite{chaudhari2017stochastic}. The inverse temperature is given by $\beta=\frac{2|\mathcal{B}|}{\Gamma}$, where $\Gamma$ is the learning rate and $\mathcal{|B|}$ is the batch size.

\begin{figure}[ht!]
\begin{center}
\includegraphics[width=17cm,height=12cm]{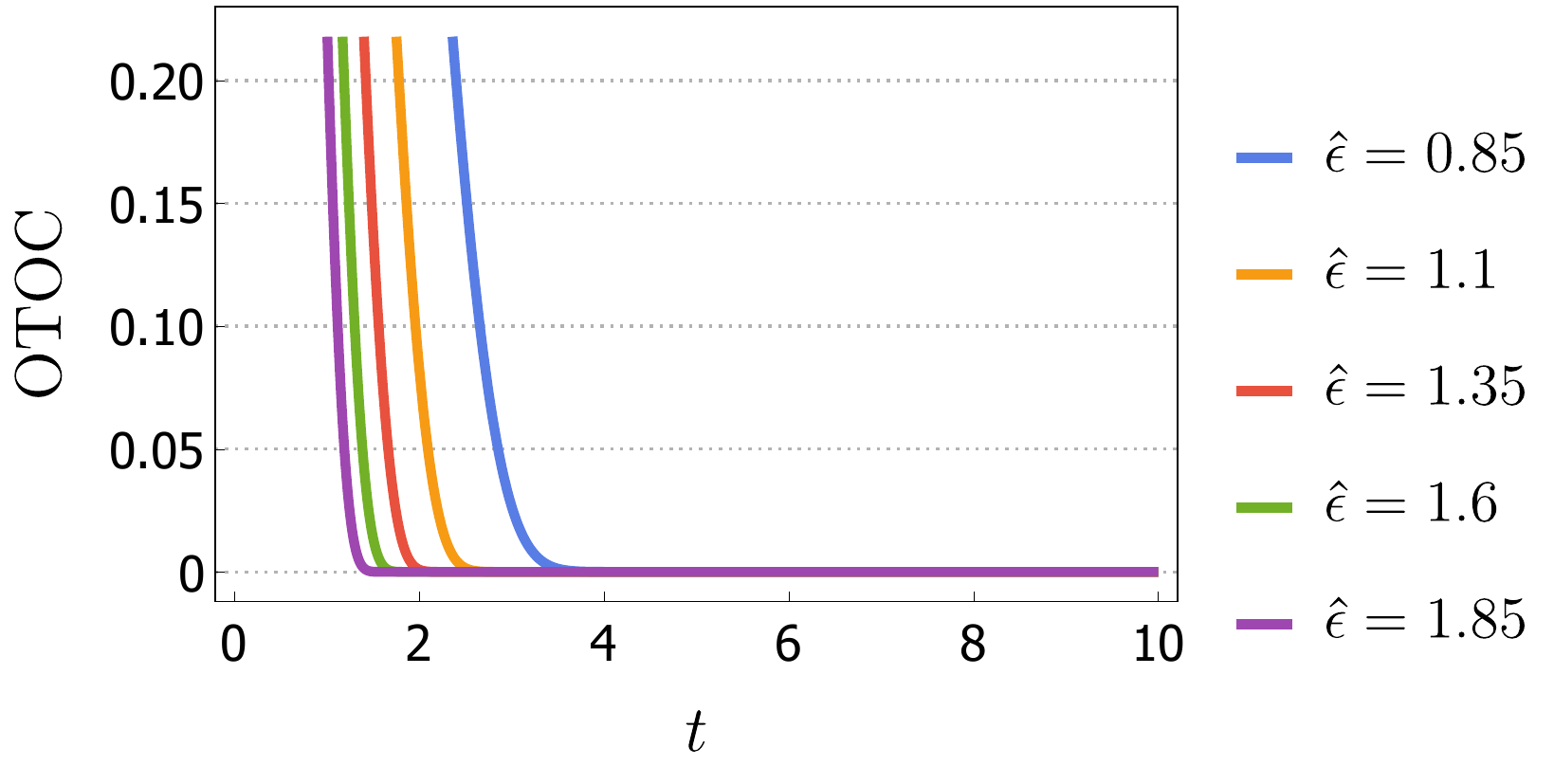}
\end{center}
\caption{Evolution of out-of-order correlator (OTOC) for a $K=2$ parameter system with a variation of eigenvalue of \textit{Encoder-Dataset} tensor $A^\infty$}
\label{fig- otoc}
\end{figure}

\begin{figure}[ht!]
\begin{center}
\includegraphics[width=17cm,height=12cm]{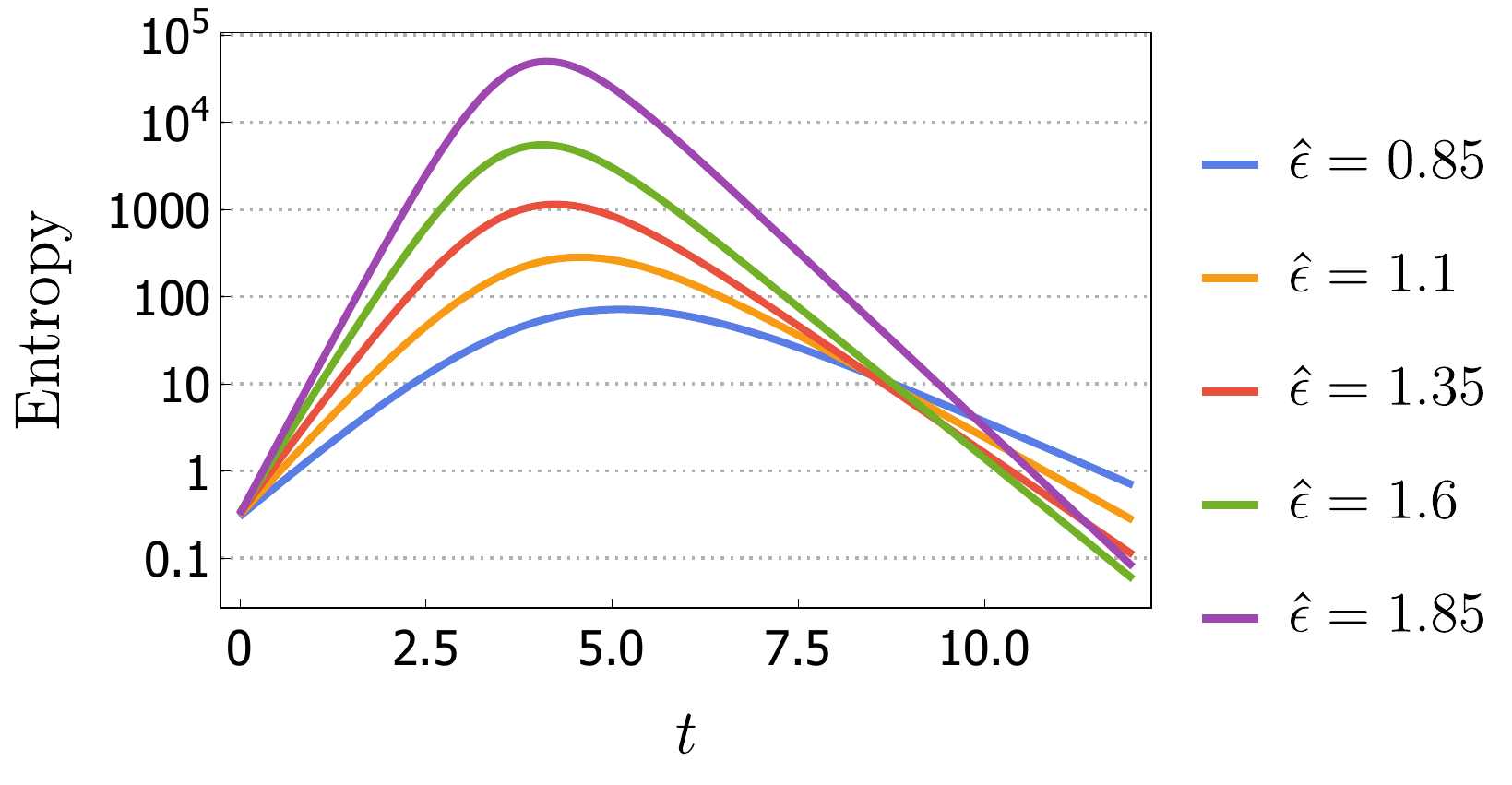}
\end{center}
\caption{Evolution of entropy for a $K=2$ parameter system with a variation of eigenvalue of \textit{Encoder-Dataset} tensor $A^\infty$}
\label{fig- entropy}
\end{figure}

\subsection{Results and Discussions}
\label{sec: results}

A $2$-dimensional independent parameter system has been studied numerically for a dataset of $N=10^3$. The evolution of parameters is governed by the equation \ref{eq- geo} while ignoring the order $\mathcal{O}(\epsilon)$ terms. The independent parameter system has the jacobian matrix $\mathcal{G}=I$. The observation matrix $B$ and $2-$rank matrix $A^\infty$ are chosen. As the dimension of the parameter system is $K=2$, the architecture or mathematically $\phi$ set runs upto $K=2$.The original $4-$rank tensor $A^\infty_{ijkl}$ is symmetric under permutation which leads to $2-$rank matrix $A^\infty_{ij}=A^\infty_{ijij}$. The observation matrix is chosen to be identity $B=I$ and the eigenvalues of the reduced matrix $A^\infty$ are taken as $\{\hat{\epsilon},\hat{\epsilon}\}$, which is varied in the results in Figures \ref{fig- logC}, \ref{fig- otoc}, and \ref{fig- entropy}. The evolution of parameters are evaluated using these inputs and then the Complexity is evaluated using equation \ref{eq- complexity reduced} with $k=2$. The evolution of Complexity is shown in Figure \ref{fig- logC}. The Lyapunov exponent is calculated by considering the change of y-axis value over the range of the x-axis value i.e. between the point of rising and point of saturation of Figure \ref{fig- logC}. The time of point of rising is shown as $t_1=0$ and the time of saturation is shown as $t_2=5$. So, mathematically, the Lyapunov exponent is given by: 
\begin{equation}
\boxed{\boxed{\lambda=\frac{\log \mathcal{C}(t)\Big|_{t=t_2}-\log \mathcal{C}(t)\Big|_{t=t_1}}{t_2-t_1}}}.
\end{equation}
Further the quantities OTOC and Entropy are plotted versus time $t$ using equations \ref{eq- otoc eva} and \ref{eq- entropy eva} in Figures \ref{fig- otoc} and \ref{fig- entropy} respectively. The scrambling time $t_*$ has been evaluated using equation \ref{eq- scramb}. 
From Figure \ref{fig- logC}, it is important to observe that as the eigenvalues of \textit{Encoder-Dataset} tensor increases, the maximum complexity of the system also increases. Thus, the role of \textit{Encoder-Dataset} matrix $A^\infty$ can be interpreted. But there can be other eigenvalue choices of matrix $A^\infty$ where the desired exponential growth and then saturation bound like behavior cannot be observed. Keeping everything unchanged, we varied the eigenvalues of the matrix $\text{eig}(A^\infty)=\{\hat\epsilon_1,\hat \epsilon_2\}$ as shown in Figure \ref{fig- complexity_dec}-\ref{fig- complexity_dec_2}. In these cases or parameter space, the relation \ref{eq- complexity reduced} and \ref{eq- otoc eva} doesn't hold. 

\begin{table}[t]
\centering
\begin{tabular}{ |p{4cm}||p{3.5cm}|p{3.5cm}|p{3.5cm}|  }
 \hline
 \multicolumn{3}{|c|}{Chaos parameters Details across Eigenvalues} \\
 \hline
 $\hat{\epsilon}$ & $\lambda$  & $t_{*}$ 
 \\
 \hline
 0.85 &  1.296  & 2.3148 
 \\
 1.1   & 1.632  &1.8382
 \\
 1.35 &  1.950  & 1.5385   
 \\
 1.6 & 2.286 & 1.3123
 \\
 1.85    &2.738 & 1.0957
 \\
 \hline
\end{tabular}
  \caption{Table showing Hyper parameter Details used in Neural Network for three different number of sampling}
  \label{table- A}
\end{table}

From the Table \ref{table- A} and Figures \ref{fig- logC}-\ref{fig- entropy}, it is evident that as the eigenvalue of the architecture increases the QNN system becomes more chaotic. The eigenvalues of the encoder-dataset matrix $A^\infty$ thus can be viewed as an essential parameter to predict the chaotic nature of the QNN even before its training. As one changes the eigenvalues of the encoder-dataset matrix, the behavior of the complexity changes.
\begin{itemize}
    \item When the eigenvalues of $A^\infty$ are all positive then the complexity shows a typical exponential growth then saturates as shown in Figure \ref{fig- logC}. In this case, all the parameters like OTOC, Lyapunov exponent can be evaluated as shown in equation \ref{eq- otoc}.
    \item If atleast one of the eigenvalue of the encoder-dataset matrix is negative, then there is no such exponential growth as shown in Figures \ref{fig- complexity_dec} and \ref{fig- complexity_dec_2}, and the relation used above i.e. \ref{eq- otoc} is not valid.
    \item On the other hand, the complexity can also have an oscillatory behavior as shown in Figure \ref{fig- complexity_osc}-\ref{fig- complexity_osc_3}, if there is atleast one complex eigenvalue, for which also the relation \ref{eq- otoc} doesn't hold. 
\end{itemize}
  It is important to note that the encoder-dataset matrix is a constant given the dataset and the architecture of the encoder circuit. So, one can always come up with an architecture for a given dataset for which the complexity of the QNN will have a typical exponential growth. The dataset for QNN is given by $\{x_i,\bar B_i\}$, so one must choose the architecture or mathematically $\phi$ set so that the matrix $A^\infty_{ij}$ is positive definite. Numerically, we computed the encoder-dataset matrix $A^\infty$ for $K=2$, based on the $\phi$ set given by 
  
  \bea \label{eq- phi ex}
  \phi_1 = \frac{1}{1-\exp(x)}+i\sin{x}, \hspace{5mm} \phi_2=\tanh{x}+i\cos{x}
  \eea 
  
  We choose a random dataset $\{x_i\}$, and evaluated the encoder-dataset matrix $A^\infty$ using the equation \ref{eq- A tensor} for the tensor $A^\infty_{ijij}$. For the $\phi$ set as shown in  \ref{eq- phi ex}, the eigenvalue of the encoder-dataset $A^\infty$ turns out to be $0.21$ and $2.53$. The higher eigenvalue dominates to show that the system will show a typical chaotic behavior with exponential growth then saturation. 
  As one chooses the $\phi$ set such that the eigenvalue of the encoder-dataset matrix is largely positive, the scrambling time decreases as shown in Table \ref{table- A}. On the other hand, the Lyapunov exponent increases. This increase in the Lyapunov exponent increases the learning rate of QNN \cite{dutta2020geometry}. Thus as one increases the eigenvalue of the architecture matrix, the learning rate of the system increases but also makes the system more chaotic. It can be intuitively argued and also known in the classical learning theory community \cite{dutta2020geometry,chaudhari2016entropysgd,chaudhari2017stochastic}, that increasing the learning rate will also decrease the probability of finding better critical points in the loss function space and thus decreases the generalization capability of the QNN. So there needs to be an optimal range of eigenvalue of the encoder-dataset matrix for which both the generalization capability of QNN and the learning rate will be optimized i.e. the QNN can find the best critical value solution in the fastest time interval possible. Finding this optimal range of the eigenvalue of the encoder-dataset matrix is beyond the scope of this paper but is an important topic that can be further looked up to.

\section{Conclusion}

The paper uses Parameterized Quantum Circuits (PQCs) in the hybrid quantum-classical framework to perform optimization of quantum data with a classical gradient-based learning algorithm like stochastic gradient descent. The optimization is executed by updates the parameters in the unitary operators of quantum circuits. The trajectory of unitaries in the unitary space is correlated with the trajectory of parameters in a Riemannian manifold called Diffusion metric \cite{Fioresi2020}. A statistical learning theory framework is introduced as a quantum analog of \cite{Bialek}. In doing so, the relation between the learning dynamics and the neural architecture of QNN is established. The relation is used to also establish the dependency of the noise in SGD on the neural architecture of QNN using the Diffusion metric. Using the definition of complexity \cite{susskind2018lectures,Brown_2016,Stanford_2014,brand2019models,Goto_2019,
Bernamonti_2020,Carmi_2017,Guo_2018,Jefferson_2017,Choudhury:2021qod,Choudhury:2020hil,Krishnan:2021faa,Khan:2018rzm,Bhattacharyya:2020iic,Bhattacharyya:2020art,Bhattacharyya:2020kgu,Bhattacharyya:2020rpy,Bhattacharyya:2019kvj,Ali:2019zcj,Ali:2018fcz,Bhattacharyya:2018bbv,Bhattacharyya:2018wym}, the paper established dependency of the parameters on complexity. The parameterized Lyapunov exponent has been derived which estimates the stability of the system.  
The paper also proves that when the system executes limit cycles or oscillations in the phase space, the generalization capability of QNN is maximized. This is consistent with the biological notion argued by \cite{YanE4185} that oscillations in phase space are important in the stability of the formation of continuous memories. The important contributions or results of the paper can be listed as follows:  
\begin{itemize}
\item Correlation between the evolution of unitary operators in the unitary space with the trajectory of parameters in the Diffusion metric. 
\item Establishing Complexity, Lyapunov exponent, OTOC, and Entropy as a function of parameters of QNN. 
\item Estimating the stability of QNN using Lyapunov exponent.
\item Proving that QNN with limit cycles or oscillations in phase space will have maximum generalization capability. 
\item Role of Encoder-dataset matrix in determining the chaotic nature of QNN before its training is established. 
\end{itemize}
 
Moreover, as neuroscience holds the fundamental architecture of neural networks,  despite the proposal of quantum processing in neurons by Fisher \cite{FISHER2015593} not much progress has been made to understand learning systems like human cognition from the perspective of quantum chaos and learning manifolds. Thus it not only becomes important to appreciate the application capability of QNN but also to analyze the quantum learning systems through the lens of statistical learning of QNN.  A possible way of connecting the human brain with the models of neuroscience is correlating the famous Hodgkin-Huxley model \cite{doi:10.1113/jphysiol.1952.sp004764} with the parameters' trajectory. Reverse engineering the QNN model that would correspond to the Quantum Hodgkin-Huxley model, can give much insight into the mechanism of the human brain.
\newpage

	\subsection*{Acknowledgements}
The research fellowship of SC is supported by the J. C. Bose National Fellowship of Sudhakar Panda. Also SC take this opportunity to thank sincerely to
Sudhakar Panda for his constant support and providing huge inspiration. SC also would line to thank School of Physical Sciences, National Institute for Science Education and Research (NISER), Bhubaneswar for providing the work friendly environment. Particularly SC want to give a separate credit to all the members of the EINSTEIN KAFFEE Berlin Alexanderplatz for providing work friendly environment, good espresso shots, delicious chocolate and caramel cakes and cookies, which helped to write the most of the part of the paper in that coffee shop in the last few months. SC also thank all the members of our newly formed virtual international non-profit consortium ``Quantum Structures of the Space-Time \& Matter" (QASTM) for elaborative discussions. SC also would like to thank all the speakers of QASTM zoominar series from different parts of the world  (For the uploaded YouTube link look at: \textcolor{red}{https://www.youtube.com/playlist?list=PLzW8AJcryManrTsG-4U4z9ip1J1dWoNgd}) for supporting my research forum by giving outstanding lectures and their valuable time during this COVID pandemic time. Last but not the least, we would like to acknowledge our debt to 
the people belonging to the various part of the world for their generous and steady support for research in natural sciences. 

\clearpage
	\appendix
\section{Detailed computation}
\label{eq- appendix}

For each datapoint $i$, the QNN produces a loss function $f_i$ at every iteration, which determines the distance between the corresponding output observation datapoint $B_i$ and the actual training datapoint $\bar{B_i}$. The loss function of the dataset $\{i\}$ is given by the average of the loss functions $f_i$ generated by the QNN across every datapoints of the dataset. We assume that the average is independent of the length of the dataset $N$ or in other words the operation will reach a thermodynamic limit with $N\rightarrow \infty$. In our approach here, we assume that the convergence of the loss function to a thermodynamic limit holds.

\bea \label{eq- loss_1}
\begin{split}
f_i=&(\bar B_i-\text{Tr}(B\mathcal{U}_\theta^\dagger\rho_{\text{in}}^i\mathcal{U}_\theta))^2\hspace{2mm}\Big[\text{from equation }\ref{eq- theta-f} \Big]\\
\Rightarrow f=&\frac{1}{N}\sum_{i=1}^N(\bar B_i-\text{Tr}(B\mathcal{U}_\theta^\dagger\rho_{\text{in}}^i\mathcal{U}_\theta))^2 \\
=&\frac{1}{N}\sum_{i=1}^N(\text{Tr}(B\mathcal{U}_{\bar\theta}^\dagger\rho_{\text{in}}^i\mathcal{U}_{\bar\theta})+\eta-\text{Tr}(B\mathcal{U}_\theta^\dagger\rho_{\text{in}}^i\mathcal{U}_\theta))^2\hspace{2mm}\Big[\text{using equation }\ref{eq- observation} \Big] \\
=&\frac{1}{N}\sum_{i=1}^N\Big\{\text{Tr}\Big(B\mathcal{U}_{\bar\theta}^\dagger\rho_{\text{in}}^i\mathcal{U}_{\bar\theta}-B\mathcal{U}_\theta^\dagger\rho_{\text{in}}^i\mathcal{U}_\theta\Big)+\eta_i\Big\}^2 \\
=&\frac{1}{N}\sum_{i=1}^N\Big\{\text{Tr}\Big(B\sum_\mu\bar\theta_\mu^*\sigma^\mu\rho_{in}^i\sum_\nu\bar\theta_\nu\sigma^\nu-B\sum_\mu\theta_\mu^*\sigma^\mu\rho_{in}^i\sum_\nu\theta_\nu\sigma^\nu\Big)+\eta_i\Big\}^2\\&\Big[\text{using equation } \ref{eq- encoder}\text{and }\ref{eq- unitary}\Big]\\
=&\frac{1}{N}\sum_{i=1}^N\Big\{-\text{Tr}\Big(\sum_{\mu,\nu}(\theta_\mu^*\theta_\nu-\bar\theta_\mu^*\bar\theta_\nu)B\sigma^\mu\rho_{\text{in}}^i\sigma^\nu\Big)+\eta_i\Big\}^2\\
=&\frac{1}{N}\sum_{i=1}^N\Big\{-\sum_{\mu,\nu}^K(\theta_\mu^*\theta_\nu-\bar\theta_\mu^*\bar\theta_\nu)\text{Tr}\Big(B\sigma^\mu\rho_{\text{in}}^i\sigma^\nu\Big)+\eta_i\Big\}^2\\&\Big[\text{using the Trace property: }\text{Tr}(\sum.)=\sum(\text{Tr}.)\Big]\\
=&\sigma_\eta^2+\frac{1}{N}\sum_{i=1}^N\sum_{\mu,\nu,\delta,\gamma}^K(\theta_\mu^*\theta_\nu-\bar\theta_\mu^*\bar\theta_\nu)(\theta_\delta^*\theta_\gamma-\bar\theta_\delta^*\bar\theta_\gamma)\text{Tr}\Big(B\sigma^\mu\rho_{\text{in}}^i\sigma^\nu\otimes B\sigma^\delta\rho_{\text{in}}^i\sigma^\gamma\Big)\\&\Big[\sum_i\eta_i^2=N\sigma_\eta^2; \sum_i \eta_i=0\text{ which leads the term with Gaussian noise $\eta$ equal to zero}\Big]\\
=&\sigma_\eta^2+\sum_{\mu,\nu,\delta,\gamma}^K(\theta_\mu^*\theta_\nu-\bar\theta_\mu^*\bar\theta_\nu)(\theta_\delta^*\theta_\gamma-\bar\theta_\delta^*\bar\theta_\gamma)\text{Tr}\Big(\underset{(a)}{\underbrace{\frac{1}{N}\sum_{i=1}^N (B\sigma^\mu\rho_{\text{in}}^i\sigma^\nu\otimes B\sigma^\delta\rho_{\text{in}}^i\sigma^\gamma}})\Big)\\
\end{split}
\eea

where $(a)$ plays the central role in determining the condition for which the loss function will reach the thermodynamic limit. The value $(a)$ is given by

\bea \label{eq- loss_2}
\begin{split}
(a)=&\frac{1}{N}\sum_{i=1}^N (B\sigma^\mu\rho_{\text{in}}^i\sigma^\nu\otimes B\sigma^\delta\rho_{\text{in}}^i\sigma^\gamma)\\
=&\frac{1}{N}\sum_{i=1}^N \Big(B\sigma^\mu\sum_{j,k}^K\phi^{*}_j(x_i)\phi_{k}(x_i)\sigma^j\ket{0}\bra{0}\sigma^k \sigma^\nu\otimes B\sigma^\delta\sum_{p,q}^K\phi^{*}_p(x_i)\phi_{q}(x_i)\sigma^p\ket{0}\bra{0}\sigma^q\sigma^\gamma\Big)\\&\Big[\text{using equations }\ref{eq- encoder}-\ref{eq- initial density}\Big]\\
=&\frac{1}{N}\sum_{i=1}^N \Big(\sum_{j,k,p,q}^K\phi^{*}_j(x_i)\phi^{*}_p(x_i)\phi_{q}(x_i)\phi_{k}(x_i)B\sigma^\mu\sigma^j\ket{0}\bra{0}\sigma^k \sigma^\nu\otimes B\sigma^\delta\sigma^p\ket{0}\bra{0}\sigma^q\sigma^\gamma\Big)\\
=&\sum_{j,k,p,q}^K\Big[\frac{1}{N}\sum_{i=1}^N \phi^{*}_j(x_i)\phi^{*}_p(x_i)\phi_{q}(x_i)\phi_{k}(x_i)\Big](B\sigma^\mu\sigma^j\ket{0}\bra{0}\sigma^k \sigma^\nu\otimes B\sigma^\delta\sigma^p\ket{0}\bra{0}\sigma^q\sigma^\gamma)\\
=&\sum_{j,k,p,q}^KA_{j'kp'q}^\infty(B\sigma^\mu\sigma^j\ket{0}\bra{0}\sigma^k \sigma^\nu\otimes B\sigma^\delta\sigma^p\ket{0}\bra{0}\sigma^q\sigma^\gamma\Big)\hspace{2mm}\Big[\text{for large }N\rightarrow \infty\Big]\\
\end{split}
\eea

Using equations \ref{eq- loss_1} and \ref{eq- loss_2}, the loss function thus can be shown as 

\bea
f=\sigma_\eta^2+\sum_{\mu,\nu,\delta,\gamma,j,k,p,q}^K A_{j'kp'q}^\infty(\theta_\mu^*\theta_\nu-\bar\theta_\mu^*\bar\theta_\nu)(\theta_\delta^*\theta_\gamma-\bar\theta_\delta^*\bar\theta_\gamma)\text{Tr}\Biggl(B\sigma^\mu\sigma^j\ket{0}\bra{0}\sigma^k \sigma^\nu\otimes B\sigma^\delta\sigma^p\ket{0}\bra{0}\sigma^q\sigma^\gamma\Biggr) \nonumber\\
\eea

Here we have introduced an encoder-dataset tensor $A^\infty$ whose elements are given by 
	\bea \label{eq- A tensor A}
A^\infty_{j'kp'q}=\lim_{N\rightarrow  \infty}\frac{1}{N}\sum_{i=1}^N \phi^{*}_j(x_i)\phi^{*}_p(x_i)\phi_{q}(x_i)\phi_{k}(x_i)	
	\eea

In equation \ref{eq- A tensor A}, we have assumed that the right hand side of the equation reaches a thermodynamic limit and an asymptotic analysis is carried out. The encoder-dataset tensor $A^\infty$ is independent of the length of the dataset and only depends on the dataset and architecture of the encoder circuit.

\newpage
\phantomsection
\addcontentsline{toc}{section}{References}
\bibliographystyle{utphys}
\bibliography{ref.bib} 

\providecommand{\href}[2]{#2}\begingroup\raggedright\begin{thebibliography}{10}

\bibitem{Arute2019}
A.~\textit{et al}, ``Quantum supremacy using a programmable superconducting
  processor,'' \href{http://dx.doi.org/10.1038/s41586-019-1666-5}{{\em Nature}
  {\bfseries 574} no.~7779, (Oct, 2019) 505--510}.
  \url{https://doi.org/10.1038/s41586-019-1666-5}.

\bibitem{Chen_2018}
Z.-Y. Chen, Q.~Zhou, C.~Xue, X.~Yang, G.-C. Guo, and G.-P. Guo, ``64-qubit
  quantum circuit simulation,''
  \href{http://dx.doi.org/10.1016/j.scib.2018.06.007}{{\em Science Bulletin}
  {\bfseries 63} no.~15, (Aug, 2018) 964–971}.
  \url{http://dx.doi.org/10.1016/j.scib.2018.06.007}.

\bibitem{commerce}
M.~Mohseni, P.~Read, H.~Neven, S.~Boixo, V.~Denchev, R.~Babbush, A.~Fowler,
  V.~Smelyanskiy, and J.~Martinis, ``Commercialize early quantum
  technologies,'' \href{http://dx.doi.org/10.1038/543171a}{{\em Nature}
  {\bfseries 543} (03, 2017) 171--174}.

\bibitem{Preskill_2018}
J.~Preskill, ``Quantum computing in the nisq era and beyond,''
  \href{http://dx.doi.org/10.22331/q-2018-08-06-79}{{\em Quantum} {\bfseries 2}
  (Aug, 2018) 79}. \url{http://dx.doi.org/10.22331/q-2018-08-06-79}.

\bibitem{Iverson_2020}
J.~K. Iverson and J.~Preskill, ``Coherence in logical quantum channels,''
  \href{http://dx.doi.org/10.1088/1367-2630/ab8e5c}{{\em New Journal of
  Physics} {\bfseries 22} no.~7, (Aug, 2020) 073066}.
  \url{http://dx.doi.org/10.1088/1367-2630/ab8e5c}.

\bibitem{Yuan1054}
X.~Yuan, ``A quantum-computing advantage for chemistry,''
  \href{http://dx.doi.org/10.1126/science.abd3880}{{\em Science} {\bfseries
  369} no.~6507, (2020) 1054--1055},
  \href{http://arxiv.org/abs/https://science.sciencemag.org/content/369/6507/1054.full.pdf}{{\ttfamily
  https://science.sciencemag.org/content/369/6507/1054.full.pdf}}.
  \url{https://science.sciencemag.org/content/369/6507/1054}.

\bibitem{Beer2020}
K.~Beer, D.~Bondarenko, T.~Farrelly, T.~J. Osborne, R.~Salzmann,
  D.~Scheiermann, and R.~Wolf, ``Training deep quantum neural networks,''
  \href{http://dx.doi.org/10.1038/s41467-020-14454-2}{{\em Nature
  Communications} {\bfseries 11} no.~1, (Feb, 2020) 808}.
  \url{https://doi.org/10.1038/s41467-020-14454-2}.

\bibitem{NIPS2016_6401}
A.~Kapoor, N.~Wiebe, and K.~Svore, ``Quantum perceptron models,'' in {\em
  Advances in Neural Information Processing Systems 29}, D.~D. Lee,
  M.~Sugiyama, U.~V. Luxburg, I.~Guyon, and R.~Garnett, eds., pp.~3999--4007.
\newblock Curran Associates, Inc., 2016.
\newblock \url{http://papers.nips.cc/paper/6401-quantum-perceptron-models.pdf}.

\bibitem{PhysRevA.100.020301}
R.~C. Wiersema and H.~J. Kappen, ``Implementing perceptron models with
  qubits,'' \href{http://dx.doi.org/10.1103/PhysRevA.100.020301}{{\em Phys.
  Rev. A} {\bfseries 100} (Aug, 2019) 020301}.
  \url{https://link.aps.org/doi/10.1103/PhysRevA.100.020301}.

\bibitem{Rebentrost_2019}
P.~Rebentrost, M.~Schuld, L.~Wossnig, F.~Petruccione, and S.~Lloyd, ``Quantum
  gradient descent and newton's method for constrained polynomial
  optimization,'' \href{http://dx.doi.org/10.1088/1367-2630/ab2a9e}{{\em New
  Journal of Physics} {\bfseries 21} no.~7, (Jul, 2019) 073023}.
  \url{https://doi.org/10.1088%2F1367-2630%2Fab2a9e}.

\bibitem{Mitarai}
K.~Mitarai, M.~Negoro, M.~Kitagawa, and K.~Fujii, ``Quantum circuit learning,''
  \href{http://dx.doi.org/10.1103/PhysRevA.98.032309}{{\em Phys. Rev. A}
  {\bfseries 98} (Sep, 2018) 032309}.
  \url{https://link.aps.org/doi/10.1103/PhysRevA.98.032309}.

\bibitem{Benedetti_2019}
M.~Benedetti, E.~Lloyd, S.~Sack, and M.~Fiorentini, ``Parameterized quantum
  circuits as machine learning models,''
  \href{http://dx.doi.org/10.1088/2058-9565/ab4eb5}{{\em Quantum Science and
  Technology} {\bfseries 4} no.~4, (Nov, 2019) 043001}.
  \url{https://doi.org/10.1088%2F2058-9565%2Fab4eb5}.

\bibitem{Bengio}
I.~Goodfellow, Y.~Bengio, and A.~Courville, {\em Deep Learning}.
\newblock The MIT Press, 2016.

\bibitem{chaudhari2016entropysgd}
P.~Chaudhari, A.~Choromanska, S.~Soatto, Y.~LeCun, C.~Baldassi, C.~Borgs,
  J.~Chayes, L.~Sagun, and R.~Zecchina, ``Entropy-sgd: Biasing gradient descent
  into wide valleys,'' 2016.

\bibitem{dutta2020geometry}
A.~Dutta and A.~Rakshit, ``Geometry perspective of estimating learning
  capability of neural networks,'' 2020.

\bibitem{pmlr-v97-goldfeld19a}
Z.~Goldfeld, E.~Van Den~Berg, K.~Greenewald, I.~Melnyk, N.~Nguyen,
  B.~Kingsbury, and Y.~Polyanskiy, ``Estimating information flow in deep neural
  networks,'' in {\em Proceedings of the 36th International Conference on
  Machine Learning}, K.~Chaudhuri and R.~Salakhutdinov, eds., vol.~97 of {\em
  Proceedings of Machine Learning Research}, pp.~2299--2308.
\newblock PMLR, Long Beach, California, USA, 09--15 jun, 2019.
\newblock \url{http://proceedings.mlr.press/v97/goldfeld19a.html}.

\bibitem{shwartzziv2017opening}
R.~Shwartz-Ziv and N.~Tishby, ``Opening the black box of deep neural networks
  via information,'' 2017.

\bibitem{fort2019emergent}
S.~Fort and S.~Ganguli, ``Emergent properties of the local geometry of neural
  loss landscapes,'' 2019.

\bibitem{10.5555/3305381.3305487}
L.~Dinh, R.~Pascanu, S.~Bengio, and Y.~Bengio, ``Sharp minima can generalize
  for deep nets,'' in {\em Proceedings of the 34th International Conference on
  Machine Learning - Volume 70}, ICML’17, p.~1019–1028.
\newblock JMLR.org, 2017.

\bibitem{lampinen2018analytic}
A.~K. Lampinen and S.~Ganguli, ``An analytic theory of generalization dynamics
  and transfer learning in deep linear networks,'' 2018.

\bibitem{chaudhari2017stochastic}
P.~Chaudhari and S.~Soatto, ``Stochastic gradient descent performs variational
  inference, converges to limit cycles for deep networks,'' 2017.

\bibitem{PhysRevLett.124.200504}
H.~Shen, P.~Zhang, Y.-Z. You, and H.~Zhai, ``Information scrambling in quantum
  neural networks,''
  \href{http://dx.doi.org/10.1103/PhysRevLett.124.200504}{{\em Phys. Rev.
  Lett.} {\bfseries 124} (May, 2020) 200504}.
  \url{https://link.aps.org/doi/10.1103/PhysRevLett.124.200504}.

\bibitem{Deutsch_2000}
D.~Deutsch and P.~Hayden, ``Information flow in entangled quantum systems,''
  \href{http://dx.doi.org/10.1098/rspa.2000.0585}{{\em Proceedings of the Royal
  Society of London. Series A: Mathematical, Physical and Engineering Sciences}
  {\bfseries 456} no.~1999, (Jul, 2000) 1759–1774}.
  \url{http://dx.doi.org/10.1098/rspa.2000.0585}.

\bibitem{YanE4185}
H.~Yan, L.~Zhao, L.~Hu, X.~Wang, E.~Wang, and J.~Wang, ``Nonequilibrium
  landscape theory of neural networks,''
  \href{http://dx.doi.org/10.1073/pnas.1310692110}{{\em Proceedings of the
  National Academy of Sciences} {\bfseries 110} no.~45, (2013) E4185--E4194},
  \href{http://arxiv.org/abs/https://www.pnas.org/content/110/45/E4185.full.pdf}{{\ttfamily
  https://www.pnas.org/content/110/45/E4185.full.pdf}}.
  \url{https://www.pnas.org/content/110/45/E4185}.

\bibitem{doi:10.1113/jphysiol.1952.sp004764}
A.~L. Hodgkin and A.~F. Huxley, ``A quantitative description of membrane
  current and its application to conduction and excitation in nerve,''
  \href{http://dx.doi.org/10.1113/jphysiol.1952.sp004764}{{\em The Journal of
  Physiology} {\bfseries 117} no.~4, (1952) 500--544},
  \href{http://arxiv.org/abs/https://physoc.onlinelibrary.wiley.com/doi/pdf/10.1113/jphysiol.1952.sp004764}{{\ttfamily
  https://physoc.onlinelibrary.wiley.com/doi/pdf/10.1113/jphysiol.1952.sp004764}}.
  \url{https://physoc.onlinelibrary.wiley.com/doi/abs/10.1113/jphysiol.1952.sp004764}.

\bibitem{KORN2003787}
H.~Korn and P.~Faure, ``Is there chaos in the brain? ii. experimental evidence
  and related models,''
  \href{http://dx.doi.org/https://doi.org/10.1016/j.crvi.2003.09.011}{{\em
  Comptes Rendus Biologies} {\bfseries 326} no.~9, (2003) 787 -- 840}.
  \url{http://www.sciencedirect.com/science/article/pii/S1631069103002002}.

\bibitem{Wang1990}
L.~P. Wang, E.~E. Pichler, and J.~Ross, ``Oscillations and chaos in neural
  networks: an exactly solvable model,''
  \href{http://dx.doi.org/10.1073/pnas.87.23.9467}{{\em Proceedings of the
  National Academy of Sciences of the United States of America} {\bfseries 87}
  no.~23, (Dec, 1990) 9467--9471}.
  \url{https://pubmed.ncbi.nlm.nih.gov/2251287}. 2251287[pmid].

\bibitem{poole2016exponential}
B.~Poole, S.~Lahiri, M.~Raghu, J.~Sohl-Dickstein, and S.~Ganguli, ``Exponential
  expressivity in deep neural networks through transient chaos,'' 2016.

\bibitem{Potapov}
A.~Potapov and M.~Ali, ``Robust chaos in neural networks,''
  \href{http://dx.doi.org/10.1016/S0375-9601(00)00726-X}{{\em Physics Letters
  A} {\bfseries 277} (12, 2000) 310--322}.

\bibitem{FISHER2015593}
M.~P. Fisher, ``Quantum cognition: The possibility of processing with nuclear
  spins in the brain,''
  \href{http://dx.doi.org/https://doi.org/10.1016/j.aop.2015.08.020}{{\em
  Annals of Physics} {\bfseries 362} (2015) 593 -- 602}.
  \url{http://www.sciencedirect.com/science/article/pii/S0003491615003243}.

\bibitem{45826}
B.~Zoph and Q.~V. Le, ``Neural architecture search with reinforcement
  learning,''
\newblock 2017.
\newblock \url{https://arxiv.org/abs/1611.01578}.

\bibitem{luo2018neural}
R.~Luo, F.~Tian, T.~Qin, E.~Chen, and T.-Y. Liu, ``Neural architecture
  optimization,'' 2018.

\bibitem{li2019random}
L.~Li and A.~Talwalkar, ``Random search and reproducibility for neural
  architecture search,'' 2019.

\bibitem{pham2018efficient}
H.~Pham, M.~Y. Guan, B.~Zoph, Q.~V. Le, and J.~Dean, ``Efficient neural
  architecture search via parameter sharing,'' 2018.

\bibitem{Fioresi2020}
S.~S. Fioresi~Rita, Chaudhari~Pratik, ``A geometric interpretation of
  stochastic gradient descent using diffusion metrics,'' {\em Entropy 22, no.
  1: 101} (2020) .

\bibitem{Brown_2016}
A.~R. Brown, D.~A. Roberts, L.~Susskind, B.~Swingle, and Y.~Zhao, ``Holographic
  complexity equals bulk action?''
  \href{http://dx.doi.org/10.1103/physrevlett.116.191301}{{\em Physical Review
  Letters} {\bfseries 116} no.~19, (May, 2016) }.
  \url{http://dx.doi.org/10.1103/PhysRevLett.116.191301}.

\bibitem{susskind2018lectures}
L.~Susskind, ``Three lectures on complexity and black holes,'' 2018.

\bibitem{CA}
A.~R. Brown, D.~A. Roberts, L.~Susskind, B.~Swingle, and Y.~Zhao, ``Complexity,
  action, and black holes,''
  \href{http://dx.doi.org/10.1103/physrevd.93.086006}{{\em Physical Review D}
  {\bfseries 93} no.~8, (Apr, 2016) }.
  \url{http://dx.doi.org/10.1103/PhysRevD.93.086006}.

\bibitem{Maldacena2016}
J.~Maldacena, S.~H. Shenker, and D.~Stanford, ``A bound on chaos,''
  \href{http://dx.doi.org/10.1007/JHEP08(2016)106}{{\em Journal of High Energy
  Physics} {\bfseries 2016} no.~8, (Aug, 2016) 106}.
  \url{https://doi.org/10.1007/JHEP08(2016)106}.

\bibitem{Shenker2014}
S.~H. Shenker and D.~Stanford, ``Black holes and the butterfly effect,''
  \href{http://dx.doi.org/10.1007/JHEP03(2014)067}{{\em Journal of High Energy
  Physics} {\bfseries 2014} no.~3, (Mar, 2014) 67}.
  \url{https://doi.org/10.1007/JHEP03(2014)067}.

\bibitem{Swingle_2017}
B.~Swingle and D.~Chowdhury, ``Slow scrambling in disordered quantum systems,''
  \href{http://dx.doi.org/10.1103/physrevb.95.060201}{{\em Physical Review B}
  {\bfseries 95} no.~6, (Feb, 2017) }.
  \url{http://dx.doi.org/10.1103/PhysRevB.95.060201}.

\bibitem{Swingle_2016}
B.~Swingle, G.~Bentsen, M.~Schleier-Smith, and P.~Hayden, ``Measuring the
  scrambling of quantum information,''
  \href{http://dx.doi.org/10.1103/physreva.94.040302}{{\em Physical Review A}
  {\bfseries 94} no.~4, (Oct, 2016) }.
  \url{http://dx.doi.org/10.1103/PhysRevA.94.040302}.

\bibitem{Swingle2018}
B.~Swingle, ``Unscrambling the physics of out-of-time-order correlators,''
  \href{http://dx.doi.org/10.1038/s41567-018-0295-5}{{\em Nature Physics}
  {\bfseries 14} no.~10, (Oct, 2018) 988--990}.
  \url{https://doi.org/10.1038/s41567-018-0295-5}.

\bibitem{Brown_2018}
A.~R. Brown and L.~Susskind, ``Second law of quantum complexity,''
  \href{http://dx.doi.org/10.1103/physrevd.97.086015}{{\em Physical Review D}
  {\bfseries 97} no.~8, (Apr, 2018) }.
  \url{http://dx.doi.org/10.1103/PhysRevD.97.086015}.

\bibitem{Bialek}
W.~Bialek, I.~Nemenman, and N.~Tishby, ``Predictability, complexity, and
  learning,'' \href{http://dx.doi.org/10.1162/089976601753195969}{{\em Neural
  Computation} {\bfseries 13} no.~11, (2001) 2409--2463},
  \href{http://arxiv.org/abs/https://doi.org/10.1162/089976601753195969}{{\ttfamily
  https://doi.org/10.1162/089976601753195969}}.
  \url{https://doi.org/10.1162/089976601753195969}.

\bibitem{2018APS..MARS28001C}
E.~{Crosson}, T.~{Jochym-O'Connor}, and J.~{Preskill}, ``{Universal quantum
  computation in thermal equilibrium},'' in {\em APS March Meeting Abstracts},
  vol.~2018 of {\em APS Meeting Abstracts}, p.~S28.001.
\newblock Jan., 2018.

\bibitem{HORNIK1991251}
K.~Hornik, ``Approximation capabilities of multilayer feedforward networks,''
  \href{http://dx.doi.org/https://doi.org/10.1016/0893-6080(91)90009-T}{{\em
  Neural Networks} {\bfseries 4} no.~2, (1991) 251 -- 257}.
  \url{http://www.sciencedirect.com/science/article/pii/089360809190009T}.

\bibitem{Altman2018}
E.~Altman, ``Many-body localization and quantum thermalization,''
  \href{http://dx.doi.org/10.1038/s41567-018-0305-7}{{\em Nature Physics}
  {\bfseries 14} no.~10, (Oct, 2018) 979--983}.
  \url{https://doi.org/10.1038/s41567-018-0305-7}.

\bibitem{Vojta_2003}
M.~Vojta, ``Quantum phase transitions,''
  \href{http://dx.doi.org/10.1088/0034-4885/66/12/r01}{{\em Reports on Progress
  in Physics} {\bfseries 66} no.~12, (Nov, 2003) 2069–2110}.
  \url{http://dx.doi.org/10.1088/0034-4885/66/12/R01}.

\bibitem{Ag_n_2019}
C.~A. Agón, M.~Headrick, and B.~Swingle, ``Subsystem complexity and
  holography,'' \href{http://dx.doi.org/10.1007/jhep02(2019)145}{{\em Journal
  of High Energy Physics} {\bfseries 2019} no.~2, (Feb, 2019) }.
  \url{http://dx.doi.org/10.1007/JHEP02(2019)145}.

\bibitem{Stanford_2014}
D.~Stanford and L.~Susskind, ``Complexity and shock wave geometries,''
  \href{http://dx.doi.org/10.1103/physrevd.90.126007}{{\em Physical Review D}
  {\bfseries 90} no.~12, (Dec, 2014) }.
  \url{http://dx.doi.org/10.1103/PhysRevD.90.126007}.

\bibitem{brand2019models}
F.~G. S.~L. Brandão, W.~Chemissany, N.~Hunter-Jones, R.~Kueng, and
  J.~Preskill, ``Models of quantum complexity growth,'' 2019.

\bibitem{Goto_2019}
K.~Goto, H.~Marrochio, R.~C. Myers, L.~Queimada, and B.~Yoshida, ``Holographic
  complexity equals which action?''
  \href{http://dx.doi.org/10.1007/jhep02(2019)160}{{\em Journal of High Energy
  Physics} {\bfseries 2019} no.~2, (Feb, 2019) }.
  \url{http://dx.doi.org/10.1007/JHEP02(2019)160}.

\bibitem{Bernamonti_2020}
A.~Bernamonti, F.~Galli, J.~Hernandez, R.~C. Myers, S.-M. Ruan, and J.~Simón,
  ``Aspects of the first law of complexity,''
  \href{http://dx.doi.org/10.1088/1751-8121/ab8e66}{{\em Journal of Physics A:
  Mathematical and Theoretical} {\bfseries 53} no.~29, (Jul, 2020) 294002}.
  \url{http://dx.doi.org/10.1088/1751-8121/ab8e66}.

\bibitem{Carmi_2017}
D.~Carmi, R.~C. Myers, and P.~Rath, ``Comments on holographic complexity,''
  \href{http://dx.doi.org/10.1007/jhep03(2017)118}{{\em Journal of High Energy
  Physics} {\bfseries 2017} no.~3, (Mar, 2017) }.
  \url{http://dx.doi.org/10.1007/JHEP03(2017)118}.

\bibitem{Guo_2018}
M.~Guo, J.~Hernandez, R.~C. Myers, and S.-M. Ruan, ``Circuit complexity for
  coherent states,'' \href{http://dx.doi.org/10.1007/jhep10(2018)011}{{\em
  Journal of High Energy Physics} {\bfseries 2018} no.~10, (Oct, 2018) }.
  \url{http://dx.doi.org/10.1007/JHEP10(2018)011}.

\bibitem{Jefferson_2017}
R.~A. Jefferson and R.~C. Myers, ``Circuit complexity in quantum field
  theory,'' \href{http://dx.doi.org/10.1007/jhep10(2017)107}{{\em Journal of
  High Energy Physics} {\bfseries 2017} no.~10, (Oct, 2017) }.
  \url{http://dx.doi.org/10.1007/JHEP10(2017)107}.

\bibitem{Bhagat:2020pcd}
K.~Y. Bhagat, B.~Bose, S.~Choudhury, S.~Chowdhury, R.~N. Das, S.~G. Dastider,
  N.~Gupta, A.~Maji, G.~D. Pasquino, and S.~Paul, ``{The Generalized OTOC from
  Supersymmetric Quantum Mechanics\textemdash{}Study of Random Fluctuations
  from Eigenstate Representation of Correlation Functions},''
  \href{http://dx.doi.org/10.3390/sym13010044}{{\em Symmetry} {\bfseries 13}
  no.~1, (2020) 44}, \href{http://arxiv.org/abs/2008.03280}{{\ttfamily
  arXiv:2008.03280 [hep-th]}}.

\bibitem{Choudhury_2020}
S.~Choudhury, ``{The Cosmological OTOC: Formulating new cosmological
  micro-canonical correlation functions for random chaotic fluctuations in
  Out-of-Equilibrium Quantum Statistical Field Theory},''
  \href{http://dx.doi.org/10.3390/sym12091527}{{\em Symmetry} {\bfseries 12}
  no.~9, (2020) 1527}, \href{http://arxiv.org/abs/2005.11750}{{\ttfamily
  arXiv:2005.11750 [hep-th]}}.

\bibitem{2017APS..DMP.T7009B}
G.~{Bentsen}, B.~{Swingle}, M.~{Schleier-Smith}, and P.~{Hayden}, ``{Measuring
  signatures of quantum chaos in strongly-interacting systems},'' in {\em APS
  Division of Atomic, Molecular and Optical Physics Meeting Abstracts},
  vol.~2017 of {\em APS Meeting Abstracts}, p.~T7.009.
\newblock Apr., 2017.

\bibitem{Bhargava:2020fhl}
P.~Bhargava, S.~Choudhury, S.~Chowdhury, A.~Mishara, S.~P. Selvam, S.~Panda,
  and G.~D. Pasquino, ``{Quantum aspects of chaos and complexity from bouncing
  cosmology: A study with two-mode single field squeezed state formalism},''
  \href{http://arxiv.org/abs/2009.03893}{{\ttfamily arXiv:2009.03893
  [hep-th]}}.

\bibitem{li2020understanding}
J.~Li, Y.~Sun, J.~Su, T.~Suzuki, and F.~Huang, ``Understanding generalization
  in deep learning via tensor methods,'' 2020.

\bibitem{zhang2016understanding}
C.~Zhang, S.~Bengio, M.~Hardt, B.~Recht, and O.~Vinyals, ``Understanding deep
  learning requires rethinking generalization,'' 2016.

\bibitem{10.5555/3295222.3295344}
B.~Neyshabur, S.~Bhojanapalli, D.~McAllester, and N.~Srebro, ``Exploring
  generalization in deep learning,'' in {\em Proceedings of the 31st
  International Conference on Neural Information Processing Systems},
  NIPS’17, p.~5949–5958.
\newblock Curran Associates Inc., Red Hook, NY, USA, 2017.

\bibitem{jiang2020generalization}
J.~Jiang, X.~Zhang, C.~Li, Y.~Zhao, and R.~Li, ``Generalization study of
  quantum neural network,'' 2020.

\bibitem{Choudhury_2021c}
S.~Choudhury, ``{The Cosmological OTOC: A New Proposal for Quantifying
  Auto-correlated Random Non-chaotic Primordial Fluctuations},''.
  \url{{https://www.preprints.org/manuscript/202102.0616/v1}}.

\bibitem{BenTov:2021jsf}
Y.~BenTov, ``{Schwinger-Keldysh path integral for the quantum harmonic
  oscillator},'' \href{http://arxiv.org/abs/2102.05029}{{\ttfamily
  arXiv:2102.05029 [hep-th]}}.

\bibitem{Lashkari_2013}
N.~Lashkari, D.~Stanford, M.~Hastings, T.~Osborne, and P.~Hayden, ``Towards the
  fast scrambling conjecture,''
  \href{http://dx.doi.org/10.1007/jhep04(2013)022}{{\em Journal of High Energy
  Physics} {\bfseries 2013} no.~4, (Apr, 2013) }.
  \url{http://dx.doi.org/10.1007/JHEP04(2013)022}.

\bibitem{Choudhury:2021qod}
S.~Choudhury, S.~P. Selvam, and K.~Shirish, ``{Circuit Complexity From
  Supersymmetric Quantum Field Theory With Morse Function},''
  \href{http://arxiv.org/abs/2101.12582}{{\ttfamily arXiv:2101.12582
  [hep-th]}}.

\bibitem{Choudhury:2020hil}
S.~Choudhury, S.~Chowdhury, N.~Gupta, A.~Mishara, S.~P. Selvam, S.~Panda, G.~D.
  Pasquino, C.~Singha, and A.~Swain, ``{Magical Chaotic Cosmological Islands:
  Generating Page Curve to solve Black Hole Information loss problem from
  Cosmological Chaos -- Complexity connection},''
  \href{http://arxiv.org/abs/2012.10234}{{\ttfamily arXiv:2012.10234
  [hep-th]}}.

\bibitem{Krishnan:2021faa}
C.~Krishnan and V.~Mohan, ``{Hints of Gravitational Ergodicity: Berry's
  Ensemble and the Universality of the Semi-Classical Page Curve},''
  \href{http://arxiv.org/abs/2102.07703}{{\ttfamily arXiv:2102.07703
  [hep-th]}}.

\bibitem{Khan:2018rzm}
R.~Khan, C.~Krishnan, and S.~Sharma, ``{Circuit Complexity in Fermionic Field
  Theory},'' \href{http://dx.doi.org/10.1103/PhysRevD.98.126001}{{\em Phys.
  Rev. D} {\bfseries 98} no.~12, (2018) 126001},
  \href{http://arxiv.org/abs/1801.07620}{{\ttfamily arXiv:1801.07620
  [hep-th]}}.

\bibitem{Bhattacharyya:2020iic}
A.~Bhattacharyya, S.~S. Haque, and E.~H. Kim, ``{Complexity from the Reduced
  Density Matrix: a new Diagnostic for Chaos},''
  \href{http://arxiv.org/abs/2011.04705}{{\ttfamily arXiv:2011.04705
  [hep-th]}}.

\bibitem{Bhattacharyya:2020art}
A.~Bhattacharyya, W.~Chemissany, S.~S. Haque, J.~Murugan, and B.~Yan, ``{The
  Multi-faceted Inverted Harmonic Oscillator: Chaos and Complexity},''
  \href{http://dx.doi.org/10.21468/SciPostPhysCore.4.1.002}{{\em SciPost Phys.
  Core} {\bfseries 4} (2021) 002},
  \href{http://arxiv.org/abs/2007.01232}{{\ttfamily arXiv:2007.01232
  [hep-th]}}.

\bibitem{Bhattacharyya:2020kgu}
A.~Bhattacharyya, S.~Das, S.~S. Haque, and B.~Underwood, ``{Rise of
  cosmological complexity: Saturation of growth and chaos},''
  \href{http://dx.doi.org/10.1103/PhysRevResearch.2.033273}{{\em Phys. Rev.
  Res.} {\bfseries 2} no.~3, (2020) 033273},
  \href{http://arxiv.org/abs/2005.10854}{{\ttfamily arXiv:2005.10854
  [hep-th]}}.

\bibitem{Bhattacharyya:2020rpy}
A.~Bhattacharyya, S.~Das, S.~Shajidul~Haque, and B.~Underwood, ``{Cosmological
  Complexity},'' \href{http://dx.doi.org/10.1103/PhysRevD.101.106020}{{\em
  Phys. Rev. D} {\bfseries 101} no.~10, (2020) 106020},
  \href{http://arxiv.org/abs/2001.08664}{{\ttfamily arXiv:2001.08664
  [hep-th]}}.

\bibitem{Bhattacharyya:2019kvj}
A.~Bhattacharyya, P.~Nandy, and A.~Sinha, ``{Renormalized Circuit
  Complexity},'' \href{http://dx.doi.org/10.1103/PhysRevLett.124.101602}{{\em
  Phys. Rev. Lett.} {\bfseries 124} no.~10, (2020) 101602},
  \href{http://arxiv.org/abs/1907.08223}{{\ttfamily arXiv:1907.08223
  [hep-th]}}.

\bibitem{Ali:2019zcj}
T.~Ali, A.~Bhattacharyya, S.~S. Haque, E.~H. Kim, N.~Moynihan, and J.~Murugan,
  ``{Chaos and Complexity in Quantum Mechanics},''
  \href{http://dx.doi.org/10.1103/PhysRevD.101.026021}{{\em Phys. Rev. D}
  {\bfseries 101} no.~2, (2020) 026021},
  \href{http://arxiv.org/abs/1905.13534}{{\ttfamily arXiv:1905.13534
  [hep-th]}}.

\bibitem{Ali:2018fcz}
T.~Ali, A.~Bhattacharyya, S.~Shajidul~Haque, E.~H. Kim, and N.~Moynihan,
  ``{Time Evolution of Complexity: A Critique of Three Methods},''
  \href{http://dx.doi.org/10.1007/JHEP04(2019)087}{{\em JHEP} {\bfseries 04}
  (2019) 087}, \href{http://arxiv.org/abs/1810.02734}{{\ttfamily
  arXiv:1810.02734 [hep-th]}}.

\bibitem{Bhattacharyya:2018bbv}
A.~Bhattacharyya, A.~Shekar, and A.~Sinha, ``{Circuit complexity in interacting
  QFTs and RG flows},'' \href{http://dx.doi.org/10.1007/JHEP10(2018)140}{{\em
  JHEP} {\bfseries 10} (2018) 140},
  \href{http://arxiv.org/abs/1808.03105}{{\ttfamily arXiv:1808.03105
  [hep-th]}}.

\bibitem{Bhattacharyya:2018wym}
A.~Bhattacharyya, P.~Caputa, S.~R. Das, N.~Kundu, M.~Miyaji, and T.~Takayanagi,
  ``{Path-Integral Complexity for Perturbed CFTs},''
  \href{http://dx.doi.org/10.1007/JHEP07(2018)086}{{\em JHEP} {\bfseries 07}
  (2018) 086}, \href{http://arxiv.org/abs/1804.01999}{{\ttfamily
  arXiv:1804.01999 [hep-th]}}.

\end{thebibliography}\endgroup



\end{document}